\documentclass[11pt]{article}
\usepackage[utf8]{inputenc}
\usepackage[T1]{fontenc}
\usepackage{lmodern}
\usepackage{amsmath,amssymb}
\usepackage[margin=2.5cm]{geometry}
\usepackage{setspace}
\usepackage{booktabs,tabularx,array,xltabular}
\usepackage{enumitem}
\usepackage[section]{placeins}
\usepackage{tikz}
\usetikzlibrary{positioning,arrows.meta}
\usepackage[numbers,sort&compress]{natbib}
\usepackage{xurl}
\usepackage{xcolor}
\usepackage[hypertexnames=false,colorlinks=true,linkcolor=blue!50!black,citecolor=blue!50!black,urlcolor=blue!50!black]{hyperref}
\setlist{itemsep=2pt,topsep=3pt}
\renewcommand{\arraystretch}{1.25}
\title{What Does a Skill Actually Do? Estimands and Evaluation Validity for Tool and Skill Use in LLM Agents: A Critical Review}
\author{Shuyang Zhang\\[4pt]\normalsize The Hong Kong Polytechnic University, Hong Kong, China}
\date{}
\hypersetup{pdftitle={What Does a Skill Actually Do? Estimands and Evaluation Validity for Tool and Skill Use in LLM Agents: A Critical Review},pdfauthor={Shuyang Zhang},pdfsubject={Critical narrative review},pdfkeywords={LLM agents, tool retrieval, agent skills, causal inference, estimand, evaluation methodology}}

\begin{document}
\maketitle

\begin{abstract}
Reported improvements from tools and reusable skills in large language model agents refer to different comparisons. This critical narrative review examines what these evaluations estimate and which conclusions their designs support. The review checks the roles of one hundred cited papers and extracts focal evaluation designs in detail from thirty-five studies. Targeted readings of thirty-five additional published or accepted studies broaden coverage of tool creation, memory, interactive benchmarks, reliability, and risk. Designs are characterized by treatment contrast, target population, outcome, budget constraint, summary measure, and identification assumptions. Analytic decompositions and counterexamples show that pairing runs on the same task does not itself identify an invocation effect when evaluation conditions on a trigger within the treated run. Paired gain and regression counts describe discordance under the coupling protocol rather than the share of tasks whose expected outcomes worsen. Total effects of deploying a module answer a different question from efficiency under a common budget. Comparisons across studies distinguish curated skill provision from retriever replacement, task populations from triggered subsets, and preparation costs from marginal usage costs. Publication status and reading depth are recorded. The review provides a methodological synthesis and a reporting checklist to help align claims about tools and skills with the comparisons their evaluation designs support.
\end{abstract}

\section{Introduction}\label{sec:1}

Tool-use methods connect LLMs to external APIs and environments~\cite{toolformer,react,toolllm,gorilla}. Benchmarks now contain thousands of callable APIs, and ToolRet combines more than 43,000 tools into a retrieval corpus~\cite{toolret}. Reusable skills add another form of capability: Agent Workflow Memory (AWM) induces routines for agent memory, whereas Agent Skill Induction (ASI) adds verified programs to the action space~\cite{awm,asi}. Selection therefore matters at several points: a retriever or router builds a shortlist, a gate decides whether a candidate is loaded, and the agent decides whether and how to follow it~\cite{toolsurvey,tabagent}. These decisions change different parts of the system and require different evaluation contrasts.

Current evaluations use several of these contrasts. ToolRet reports both retrieval quality and downstream success after replacing the retriever~\cite{toolret}. AWM and ASI report gains from deploying their respective adaptation packages~\cite{awm,asi}. AI Agents That Matter shows why accuracy must also be compared with cost and simple resampling baselines when assessing efficiency~\cite{agentsmatter}. Recent skill studies add more specific questions. Cho and Park report a positive retrieved-versus-skipped lift together with a negative paired contrast on tasks where retrieval occurred~\cite{skillfollowing}. Hajimiri et al.\ report that online augmentation gains can disappear against a vanilla web actor given more steps within a comparable budget~\cite{worththeirtokens}. Tank and Nama separate paired gains from regressions~\cite{regressiontax}, while SkillsBench compares curated bundles with no skills~\cite{skillsbench}. These findings can coexist because they answer different questions. These frontier findings motivate design-specific analysis; they do not establish how frequently any failure occurs across agents.

Tool learning~\cite{toolsurvey}, agent evaluation and benchmarking~\cite{yehudai2026,mohammadi2025,nageshwaran2026,kehkashan2026}, and trajectory-level failure attribution~\cite{wang2026tse} have all been surveyed. These surveys organize the literature by capability, benchmark, metric, or failure type; they also discuss validity, cost, reproducibility, and deployment interpretation~\cite{nageshwaran2026,kehkashan2026}. A directly related methodological study by Huang asks what an LLM-agent leaderboard rank compares~\cite{huang2026leaderboard}. It specifies the target population, measurement source, resource rule, common support, and uncertainty needed to interpret pairwise system comparisons. Estimand-based reasoning is therefore already part of agent evaluation. This review focuses on comparisons that enable, replace, or trigger a tool or skill component within an agent. The analysis examines how conditioning on a within-run trigger interacts with the coupling of stochastic runs, and why paired outcome flips differ from the share of tasks whose expected outcomes worsen. These component-level questions complement Huang's analysis of system rankings.

This review analyzes these distinctions explicitly and synthesizes them across studies, without assuming that the source authors overlooked the limitations. Skill Following, for example, already describes its retrieval-invoked contrast as protocol-conditional~\cite{skillfollowing}; Section~\ref{sec:3.5} makes the coupling and selection terms explicit. This review contributes:

\begin{enumerate}
\item a framework that separates measurement levels from causal paths and describes evaluation designs along six axes, stated in potential-outcome notation (Section~\ref{sec:3});
\item an analysis of recurring interpretations, using analytic counterexamples to examine trigger-conditioned paired contrasts, the interpretation of stratum effects as invocation effects, and paired gain and regression counts and their naive task-level summaries; this analysis also distinguishes total effects from budget-constrained effects (Sections~\ref{sec:3.4}--\ref{sec:3.8});
\item study-level evidence tables for thirty-five studies examined through full-text extraction of the focal evaluation designs (Section~\ref{sec:4} and Appendix~\ref{app:matrix}), followed by a synthesis organized by measurement level (Section~\ref{sec:4});
\item a reporting checklist that links each item to the applicable designs, and a set of open problems (Section~\ref{sec:5}).
\end{enumerate}

\section{Scope and Review Method}\label{sec:2}

This critical narrative review examines the design logic of representative studies. Effect sizes are not pooled, and study quality is not graded. The search covered arXiv, the ACL Anthology, and conference proceedings using combinations of \emph{tool retrieval}, \emph{tool selection}, \emph{skill retrieval}, \emph{skill routing}, \emph{skill library}, \emph{counterfactual}, and \emph{evaluation}. Citations were tracked backward and forward from studies of actual skill use~\cite{skillfollowing} and component replacement~\cite{tabagent}. Selection was purposive and focused on contrasts that differ in treatment, population, pairing, or budget, including results in which skills help and results in which they do not. Evaluation-protocol studies were included when they bear directly on these comparisons. Publisher records, the ACL Anthology, PMLR, OpenReview records, and author-linked texts were consulted to check publication status and identify published evidence for retrieval, workflow memory, executable skills, cost, and evaluator stability, as well as related methodological work such as Huang's study~\cite{huang2026leaderboard}.

The role of every cited work was checked against its use in the manuscript: one hundred papers, one book, and one guideline. Thirty-five papers received detailed extraction of the focal evaluation designs used in the synthesis. These comprise five published works establishing the comparison structure, thirteen frontier case studies (Tables~\ref{tab:published} and~\ref{tab:matrix}), and seventeen studies covering retrieval policies, module removal, memory horizons, execution costs, and reliability or risk protocols. Selection depended on whether a study's empirical comparisons, ablations, or reliability and risk evaluations directly supported a central methodological distinction. Studies used primarily to define an architecture, benchmark population, or endpoint received targeted reading. This purposive distinction concerns the evidence needed for the review's claims, not study quality or whether a result was favorable.

For the thirty-five detailed studies, the extraction recorded the focal arms, unit, population and inclusion rules, runs and coupling, conditioning events, budget, endpoint, source version, and supporting locations (Appendix~\ref{app:matrix}). Extraction concerns the comparisons cited in this review, rather than every experiment in each paper. Unreported details are identified relative to the cited sections; adaptive attempts, validation folds, candidate samples, human ratings, and repeated grading are distinguished from independent agent executions. The role check across all cited works drew on the detailed extractions and targeted readings; it did not involve full-text extraction of every cited paper.

Targeted readings of thirty-five further papers cover methods and evaluation and are organized by topic in Appendix~\ref{app:coverage}. The detailed and targeted groups are disjoint and together contain seventy studies. The other thirty papers comprise nineteen methodological foundations, seven surveys or related-work positioning sources, and four technical-background sources. The book and guideline provide additional conceptual foundations. Huang's methodological paper was read in full for positioning, outside the empirical design table. These roles describe how sources are used here; papers used for methodological or background support may themselves contain experiments.

Official venue lists, OpenReview records, proceedings, and publisher records were used to verify acceptance or publication, with the version actually read recorded separately. Preprint and proceedings versions of the same work count as one study. Methodological sources cover post-treatment selection, counterfactual distributions, simulation coupling, retrieval metrics, temporal abstraction, statistical uncertainty, and sequential off-policy evaluation. These sources provide analytical background; they are outside both empirical evidence groups.\footnote{For Montgomery et al., the reading source was the author-hosted manuscript at \url{https://cpb-us-e1.wpmucdn.com/sites.dartmouth.edu/dist/5/2293/files/2021/03/post-treatment-bias.pdf}; equivalence to the final typeset text was not verified.} Numbers are quoted as reported and are not pooled across benchmarks, agents, or budgets. One author selected and characterized the studies.

Table~\ref{tab:coverage} maps the topical literature to six aspects of tool and skill evaluation. The coverage deliberately extends beyond skill retrieval: constructing a tool, learning when to call it, transferring experience, and measuring an interactive trajectory change different parts of an evaluation. Statistical and causal-inference references supply the analytical foundation separately. This map documents the scope of the selected literature, not an exhaustive search or a quality ranking.

\begin{table}[!htbp]
\centering\footnotesize
\caption{Coverage of the topical literature and its role in the review}\label{tab:coverage}
\begin{tabularx}{\linewidth}{>{\raggedright\arraybackslash}p{3.0cm}>{\raggedright\arraybackslash}p{3.0cm}>{\raggedright\arraybackslash}X}
\toprule
\textbf{Aspect} & \textbf{Sources} & \textbf{Evaluation question} \\
\midrule
Tool learning, creation, and retrieval & \cite{toolformer,react,toolllm,gorilla,toolret,reinvoke,toolkengpt,creator,craft,latm} & Is the comparison about tool availability, learned calling, generated tools, or replacement of a retrieval component? \\
Skill and memory construction & \cite{awm,asi,voyager,reflexion,expel} & Is the object a fixed library, within-task adaptation, or transfer of experience across tasks? \\
Benchmarks, outcomes, and protocols & \cite{appworld,stabletoolbench,swebench,webarena,mind2web,workarena,apibank,toolsandbox,taubench,agentboard,agentbench,toolemu,bfcl,osworld} & Which tasks, histories, environments, success rules, and risks does the score represent? \\
Frontier component and trajectory studies & \cite{skillfollowing,worththeirtokens,regressiontax,skillsbench,skillapt,radeg,shadowing,confgated,replaygap,vasudev2026,menu,capabilitypages,tabagent,betterturns} & What do trigger conditioning, paired flips, gates, and replay establish in the reported protocols? \\
Evaluation methodology & \cite{agentsmatter,huang2026leaderboard} & Which resource constraints and target populations make system comparisons interpretable? \\
Related reviews & \cite{toolsurvey,yehudai2026,mohammadi2025,nageshwaran2026,kehkashan2026,wang2026tse} & How does this component-level analysis relate to tool-learning, benchmarking, and trajectory-analysis syntheses? \\
\bottomrule
\end{tabularx}
\end{table}

\paragraph{Publication status and evidential role.} Publication status is reported separately from design validity. Published means a journal or proceedings record was located. Accepted means that an official venue list or OpenReview venue record confirms acceptance; main-conference, Findings, and workshop venues are distinguished. Preprint means that acceptance was not verified in this check; it does not establish that the work has not been accepted. Of the thirteen frontier cases, five have verified acceptance and eight remain preprints under this definition. These cases illustrate particular protocols and motivate hypotheses, not prevalence claims or general effect sizes. Peer review does not itself establish identification or independent replication. The algebraic results and counterexamples in Section~\ref{sec:3} follow from their stated assumptions independently of the frontier findings.

\paragraph{Use of generative AI.} Generative AI tools were used to assist with literature retrieval, drafting, reference compilation, and manuscript preparation. The author takes responsibility for the manuscript and its interpretation of the cited sources.

\section{A Framework for Tool and Skill Evaluation}\label{sec:3}

\subsection{Measurement Levels and Causal Paths}\label{sec:3.1}

The framework distinguishes five measurement levels in tool and skill evaluation:

\begin{itemize}
\item \textbf{Exposure:} which candidates the agent can see or call, such as a shortlist, a tool menu, or skill names and descriptions preloaded into the system prompt~\cite{menu,toolret}.
\item \textbf{Retrieval:} which candidates a retriever or router ranks highly for the current query or state~\cite{toolret,reinvoke}.
\item \textbf{Activation:} whether a surfaced candidate is loaded into context or executed~\cite{skillapt,radeg}.
\item \textbf{Use:} whether the agent follows the loaded procedure or tool output, and how~\cite{skillfollowing,asi}.
\item \textbf{Outcome:} task success, together with cost, latency, and risk.
\end{itemize}

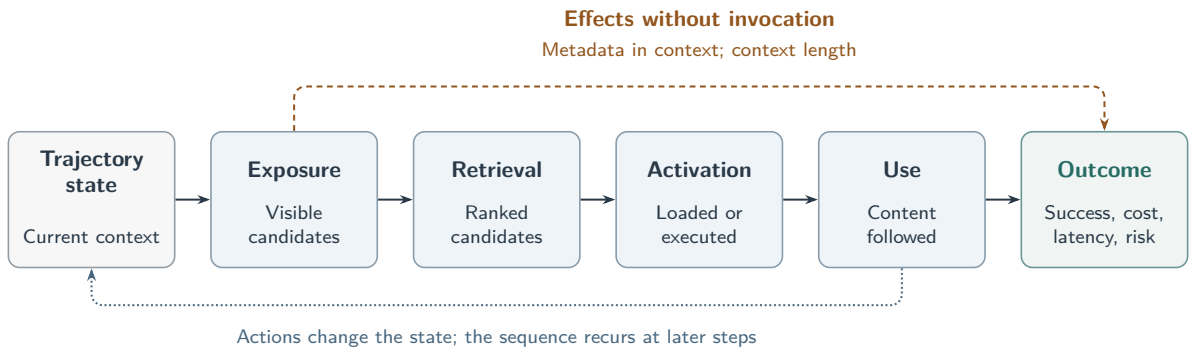
\begin{figure}[t]
\centering
\begingroup
\definecolor{pathInk}{HTML}{293947}
\definecolor{pathBlue}{HTML}{48677F}
\definecolor{pathFill}{HTML}{EFF4F8}
\definecolor{pathTeal}{HTML}{276B65}
\definecolor{pathAmber}{HTML}{92551D}
\begin{tikzpicture}[x=1mm,y=1mm,
  every node/.style={font=\sffamily\fontsize{8.5}{10}\selectfont,
    text=pathInk,align=center,inner sep=0pt},
  box/.style={draw=pathBlue!65,fill=pathFill,line width=0.6pt,
    rounded corners=1.5mm,minimum width=22mm,minimum height=18mm},
  title/.style={anchor=base,font=\sffamily\bfseries\fontsize{8.5}{10}\selectfont},
  detail/.style={anchor=base,font=\sffamily\fontsize{7.4}{9}\selectfont},
  arr/.style={-{Stealth[length=1.7mm,width=1.2mm]},line width=0.75pt,draw=pathInk}]
\node[box,draw=pathInk!50,fill=black!3] (st) at (11,0) {};
\node[box] (ex) at (37.8,0) {};
\node[box] (re) at (64.6,0) {};
\node[box] (ac) at (91.4,0) {};
\node[box] (us) at (118.2,0) {};
\node[box,draw=pathTeal!70,fill=pathTeal!7] (ou) at (145,0) {};
\node[title] at (11,4.5) {Trajectory};
\node[title] at (11,1.0) {state};
\node[detail] at (11,-5.8) {Current context};
\node[title] at (37.8,3.0) {Exposure};
\node[detail] at (37.8,-2.6) {Visible};
\node[detail] at (37.8,-5.8) {candidates};
\node[title] at (64.6,3.0) {Retrieval};
\node[detail] at (64.6,-2.6) {Ranked};
\node[detail] at (64.6,-5.8) {candidates};
\node[title] at (91.4,3.0) {Activation};
\node[detail] at (91.4,-2.6) {Loaded or};
\node[detail] at (91.4,-5.8) {executed};
\node[title] at (118.2,3.0) {Use};
\node[detail] at (118.2,-2.6) {Content};
\node[detail] at (118.2,-5.8) {followed};
\node[title,text=pathTeal] at (145,3.0) {Outcome};
\node[detail] at (145,-2.6) {Success, cost,};
\node[detail] at (145,-5.8) {latency, risk};
\draw[arr] (st.east) -- (ex.west);
\draw[arr] (ex.east) -- (re.west);
\draw[arr] (re.east) -- (ac.west);
\draw[arr] (ac.east) -- (us.west);
\draw[arr] (us.east) -- (ou.west);
\draw[arr,draw=pathAmber,dash pattern=on 2.2pt off 1.7pt,
  rounded corners=1.5mm] (ex.north) -- (37.8,15) -- (145,15) -- (ou.north);
\node[title,text=pathAmber] at (91.4,23) {Effects without invocation};
\node[detail,text=pathAmber] at (91.4,19) {Metadata in context; context length};
\draw[arr,draw=pathBlue,densely dotted,rounded corners=1.5mm]
  (us.south) -- (118.2,-14) -- (11,-14) -- (st.south);
\node[detail,text=pathBlue] at (64.6,-19)
  {Actions change the state; the sequence recurs at later steps};
\end{tikzpicture}
\endgroup
\caption{Measurement levels in tool and skill use (solid arrows) and two paths that a purely sequential reading omits: a direct effect of exposure on the outcome (dashed) and feedback from actions to the trajectory state (dotted). The figure is a reading aid, not a complete causal graph.}\label{fig:paths}
\end{figure}

These levels describe a typical workflow and clarify what an evaluation measures. They are not a complete causal graph. The order can vary: exposure may precede retrieval (preloaded metadata) or follow it (a retrieved shortlist), and in multi-step agents, the sequence can recur at every step. Each decision changes the state on which later decisions depend. A skill can also affect the outcome without passing through every level. Skill names and descriptions in the system prompt can change behavior when the skill is never invoked~\cite{regressiontax}, and a larger context can change behavior even when selection is correct~\cite{shadowing}. Figure~\ref{fig:paths} shows the typical order together with this direct path and the state feedback.

Precise event definitions matter because several studies condition on them. In this review, a candidate is \emph{surfaced} when it appears in the agent's visible candidate set; \emph{retrieval is called} ($R^{\mathrm{call}}=1$) when the agent or harness issues a retrieval request; \emph{retrieval returns} ($R^{\mathrm{ret}}=1$) when that request yields at least one candidate; a skill is \emph{loaded} ($R^{\mathrm{load}}=1$) when its body enters the context; a tool is \emph{executed} when a call is issued and returns; and content is \emph{followed} when the agent's subsequent actions depend on it according to the study's definition, for example through a behavioral annotation~\cite{skillfollowing}. A call that returns an empty set differs from one that returns candidates, and the trigger rate and the triggered group change with the event chosen. Studies differ in these definitions; the definition used is noted whenever a result depends on it. Below, $R$ denotes whichever of these events a study conditions on.

Executable skills that bundle several primitive actions also have a useful connection to temporal abstraction. In the options framework, an option specifies an initiation set, an internal policy, and a termination condition; one option execution may span several primitive steps~\cite{sutton1999}. For the comparisons reviewed here, this motivates reporting the action granularity and internal execution cost when a skill call counts as a single high-level step.

\subsection{Notation, Units, and Coupling}\label{sec:3.2}

The analysis uses the potential-outcomes framework~\cite{rubin1974,holland1986,imbensrubin2015}. Let $x$ be a task drawn from a task population, and let $a$ denote a configuration of the agent, such as an enabled library, a particular retriever, or forced loading of a skill. Agent runs are stochastic even at fixed $x$, because of sampling, serving nondeterminism, and environment variation. Here, $Y_a$ denotes the outcome of one run under configuration $a$, and $\mu_a(x)=\mathbb{E}[Y_a\mid x]$ denotes its expectation over runs. When a module is available, $R_a\in\{0,1\}$ records whether retrieval is triggered in that run. $R$ is a post-treatment variable. It exists only in arms where the module is available, and it may depend on the same random events that determine $Y_a$.

Three choices must be fixed before any comparison is meaningful. The first is the \emph{unit}: a task, a task paired with a random seed, or a pair of runs. The second is the \emph{coupling} of arms. Arms can run independently given the task, share a random seed as in several studies below, or branch from a common recorded state. A shared seed specifies a joint sampling protocol. Common-random-number simulation likewise distinguishes shared random inputs from their effect on the variance of a comparison: the covariance of the outputs depends on how the shared inputs generate them~\cite{glasserman1992}. The dependence induced by the shared seed should be described, including whether environment and serving randomness are also controlled; a coupling does not require the two arms to produce similar trajectories. The third is the \emph{number of runs} per task and arm. It determines how precisely $\mu_a(x)$ can be estimated and which inferences are possible. A single run gives an unbiased but very imprecise estimate.

The clinical-trial addendum ICH~E9(R1) requires every estimand to specify five attributes: the population, the treatment, the outcome variable, the handling of intercurrent events, and the population-level summary~\cite{ich2019}. The same structure applies to agent components. Intercurrent events here include not triggering retrieval, retrieving a skill without loading it, loading it and ignoring it, infrastructure failures, and exhausting the step or token budget. Much of the ambiguity reviewed below comes from leaving one of these attributes, or the unit and coupling, unstated.

\subsection{Six Axes for Describing a Design}\label{sec:3.3}

The framework describes each design along six axes rather than using a single list of ``effects'' (Table~\ref{tab:axes}). The treatment contrast specifies what changes. It can be the availability of a module, a swap of one component for another (a retriever, a shortlist head, a tool menu), an activation policy, or the composition of a library. The target population specifies the tasks or states over which the quantity is averaged: all tasks, a subgroup defined before treatment, a subgroup defined by a post-treatment event such as triggering, or a set of states within trajectories. The outcome may be success, cost, or risk exposure. The budget constraint may be absent, an ex-ante cap, or an attempt to match realized spending. The summary measure may be a difference in means, a count of paired discordances, the share of tasks whose expected outcome falls, or a value function. The identification assumptions specify the conditions under which the summary has a causal interpretation rather than describing one protocol.

\begin{table}[!htbp]
\centering\small
\caption{Six axes for describing a tool or skill evaluation design}\label{tab:axes}
\begin{tabularx}{\linewidth}{>{\raggedright\arraybackslash}p{2.6cm}>{\raggedright\arraybackslash}X}
\toprule
\textbf{Axis} & \textbf{Values found in the literature} \\
\midrule
Treatment contrast & Availability (module on vs.\ off); component swap (retriever, shortlist head, menu constructor); activation policy (load vs.\ abstain, gate on vs.\ off); library composition (size, content, admission rule) \\
Target population & All tasks; pre-treatment subgroup (domain, difficulty tier); post-treatment subgroup (tasks where retrieval was triggered); states within trajectories \\
Outcome & Success or reward; tokens, calls, latency; exposure to risky candidates; intermediate failures \\
Budget constraint & None; ex-ante cap with a stated allocation rule; approximate matching of realized spending \\
Summary measure & Difference in means; paired discordance counts; share of tasks whose expected outcome falls; state-level value contrast \\
Identification & Randomized or paired runs over a task distribution; stated coupling between arms; principal ignorability for post-treatment strata; sequential ignorability and positivity for state-level contrasts from logs \\
\bottomrule
\end{tabularx}
\end{table}

\subsection{Commonly Reported Quantities and Their Limits}\label{sec:3.4}

Table~\ref{tab:quantities} organizes commonly reported quantities along these axes. Two of them are descriptive, not causal. \emph{Component metrics} such as Recall@k, nDCG, or Hit@1 describe a retriever on a labeled query pool. Normalized discounted cumulative gain combines graded relevance judgments with rank discounts and comparison to an ideal ordering~\cite{jarvelin2002}. Thus, the relevance labels and query population are part of its interpretation; downstream agent success requires a separate outcome evaluation. \emph{Retrieved-versus-skipped contrasts}, $\mathbb{E}[Y_1\mid R_1=1]-\mathbb{E}[Y_1\mid R_1=0]$ or variants that mix arms, compare two groups that the agent itself selected. Both are precisely definable statistical quantities, but neither measures a causal effect. Skill Following measures the difficulty gap between the two groups with the paired skill-disabled runs and finds it substantial~\cite{skillfollowing}.

The \emph{total effect} of a contrast between configurations $a$ and $a'$ over the task population is
\begin{equation}\label{eq:total}
\tau_{a,a'}=\mathbb{E}_x\bigl[\mu_a(x)-\mu_{a'}(x)\bigr].
\end{equation}
With availability as the contrast, $\tau$ answers ``should this module, as specified, be switched on?''. With a component swap, it answers ``should component $A$ replace component $B$?''. Both are identified by running both arms on the same tasks, and both are the same kind of quantity. Most retrieval papers that report end-to-end gains estimate a component-swap total effect: the effect of better retrieval, not the effect of skills relative to none. TabAgent's replacement of an LLM shortlist head by a classifier, which reportedly maintains task-level success on AppWorld~\cite{tabagent,appworld}, is also a component-swap comparison, with shortlisting cost as a second outcome.

\begin{table*}[!htbp]
\centering\footnotesize
\caption{Commonly reported quantities in tool and skill evaluation, what each supports, and what it does not support}\label{tab:quantities}
\begin{tabularx}{\linewidth}{>{\raggedright\arraybackslash}p{2.6cm}>{\raggedright\arraybackslash}p{3.0cm}>{\raggedright\arraybackslash}X>{\raggedright\arraybackslash}X>{\raggedright\arraybackslash}p{1.9cm}}
\toprule
\textbf{Quantity} & \textbf{Contrast; population} & \textbf{Supports} & \textbf{Does not support} & \textbf{Examples} \\
\midrule
Component metric (Recall@k, nDCG) & None; labeled query pool & Ranking quality on that pool & Any effect on agent outcomes; transfer to other query sources & \cite{toolret,reinvoke} \\
Retrieved vs.\ skipped contrast & None; post-treatment groups & Description of where the agent retrieves & Any causal effect; groups differ in difficulty & criticized in \cite{skillfollowing} \\
Total availability effect & Module on vs.\ off; all tasks & Deploying the module as specified & Resource efficiency; which path produced the effect & \cite{skillsbench,regressiontax,vasudev2026} \\
Component-swap effect & Component $A$ vs.\ $B$; all tasks, or fixed evidence paths & Replacing $B$ by $A$ in that agent & Effect of skills vs.\ no skills; value of an additional action & \cite{menu,capabilitypages,tabagent,confgated} \\
Budget-constrained comparison & Configurations under a common cap; all tasks & Choice between configurations at that budget & Total effect without the cap; exactness if the cap is approximate & \cite{worththeirtokens} \\
Composition effect & Library versions; all tasks & Effect of growing or curating the library & Effects of individual skills; later library states & \cite{shadowing,awm,asi} \\
Trigger-conditioned paired contrast & Availability; tasks where the treated run triggered & Protocol-specific diagnostic of the triggered segment & A run-level invocation effect without coupling assumptions (Section~\ref{sec:3.5}) & \cite{skillfollowing} \\
Paired discordance counts & Any; all tasks & Frequency of flips under the protocol & Share of tasks harmed; individual harm (Section~\ref{sec:3.7}) & \cite{regressiontax,vasudev2026} \\
Task-level degradation share & Any; all tasks & Share of tasks whose expected outcome falls, reported as confidently and possibly degraded shares & Mechanism of the fall; naive plug-in estimates from few runs & none found \\
State-level intervention value & Load vs.\ abstain at a state; states & Activation decisions under a fixed continuation policy & The deployed gate's total effect (Section~\ref{sec:3.8}) & \cite{skillapt} \\
Execution-selection value & Run vs.\ skip the downstream agent; query--bundle pairs & Deciding whether to execute, given a utility for skipped runs & Effect of loading a skill within a run & \cite{radeg} \\
\bottomrule
\end{tabularx}
\end{table*}

\paragraph{Budget constraints.} A module that consumes tokens changes two things at once: what the agent knows and how much computation it spends. If the treatment is ``deploy this module with its overhead'', a well-designed comparison with and without the module identifies the total effect of that package. The extra computation is part of the treatment and one path of its effect, not a confounder. Separating an overall effect from effects transmitted through particular intermediate variables is the concern of causal mediation analysis~\cite{pearl2001}; a budget restriction changes the comparison and does not itself identify such a path-specific effect. A budget-constrained comparison answers a different question: which configuration to choose when total spending is capped. Defining it requires an ex-ante cap $B$ and a rule for allocating $B$ within each arm, because realized spending is itself a post-treatment outcome. The frontier web-agent study of Hajimiri et al.\ illustrates an approximate resource comparison~\cite{worththeirtokens}. Its control arm extends the vanilla actor's horizon from 10 to 15 steps and adds accessibility-tree pruning. The authors state that this step cap approximates, and does not exactly match, the token budgets of the augmented methods; in most tasks the control spends fewer tokens, and in a few it spends slightly more. The finding that augmentation gains often vanish therefore concerns this approximate budget constraint. It does not show that comparisons without budget matching are invalid. A complete report gives the total effect, the costs in each arm, and, where resource efficiency is claimed, a budget-constrained comparison or a cost--success frontier~\cite{agentsmatter}.

\subsection{Trigger-Conditioned Paired Contrasts}\label{sec:3.5}

A natural remedy for the retrieved-versus-skipped contrast is to pair each task with a run in which the module is unavailable, and then restrict the comparison to tasks where the module arm triggered retrieval. Skill Following formalizes this as the Retrieval-Invoked Actual-Use Effect (RAE), using the returned-skill event $R^{\mathrm{ret}}$ as the trigger~\cite{skillfollowing}. Pairing removes the between-task difficulty gap that makes the retrieved-versus-skipped contrast misleading. Taken alone, it does not make the restricted contrast a causal effect of invocation. More generally, selecting observations using a variable affected by treatment can bias an experimental comparison even when the original treatment assignment was randomized~\cite{montgomery2018}.

Consider a protocol that produces, for each task, one run in the module arm with outcome $Y_1$ and trigger $R_1$, and one run in the control arm with outcome $Y_0$. The two runs may be coupled in any way that preserves their marginal distributions: they may be independent given $x$, share a random seed, or branch from a common prefix. A common prefix is a coupling of the same availability contrast only if it leaves the defined marginal distribution of each arm unchanged; branching from a recorded mid-run state more naturally defines the state-level contrast of Section~\ref{sec:3.8}. For any coupling that preserves the marginals, the population value of the paired statistic decomposes as
\begin{equation}\label{eq:pair}
\begin{aligned}
D_{\mathrm{pair}}=\mathbb{E}\bigl[Y_1-Y_0\mid R_1=1\bigr]
={}&\underbrace{\mathbb{E}\bigl[\mu_1(x)-\mu_0(x)\mid R_1=1\bigr]}_{\text{propensity-weighted task effect}}\\
&+\underbrace{\mathbb{E}\bigl[Y_1-\mu_1(x)\mid R_1=1\bigr]}_{\text{treated-run selection}}
-\underbrace{\mathbb{E}\bigl[Y_0-\mu_0(x)\mid R_1=1\bigr]}_{\text{control-run selection}}.
\end{aligned}
\end{equation}
The first term averages task-level total effects with weights proportional to each task's trigger probability. It is still an availability effect, not the effect of invoking a skill. The second term is non-zero whenever triggering is correlated with the treated run's own outcome, for example because an agent retrieves more often after an early mistake. The third term depends on the coupling. If the control run is independent of the treated run given $x$, the third term vanishes and Equation~\eqref{eq:pair} reduces to the first two terms. Under a shared seed or another dependent coupling, the third term can be non-zero and can offset the second term in part or in full.

\paragraph{An analytic counterexample.} Suppose all tasks are identical and the module has no effect on the distribution of outcomes: in each arm a run succeeds with probability $0.5$, represented by a fair coin $U$. Let the module arm trigger retrieval exactly when its coin shows $U=1$. The true effect is zero for every task. With an independent control run, the paired statistic has expectation $1-0.5=0.5$, entirely due to the treated-run selection term. If instead both arms read the same coin, the two selection terms cancel and the paired statistic is $0$. The example shows that pairing on the task does not remove selection that occurs within the task, and that the size of the resulting bias depends on the coupling. It does not show the direction or size of the bias under any particular shared-seed protocol.

Skill Following runs both arms with the same task prompt, generation seed, and decoding configuration (its Section~4.3). This defines a dependent coupling. The paper does not specify how the shared seed links the random events driving retrieval in the skill-enabled run to those driving success in the skill-disabled run, whose prompt lacks the tool definition. Both selection terms in Equation~\eqref{eq:pair} are therefore unknown, and so is their difference. The authors themselves describe RAE as ``a protocol-conditional paired outcome signal'', state that it is not an unbiased causal effect over a pre-treatment task population, and list the further ablations that a finer causal decomposition would need, including metadata-only retrieval and oracle retrieval (Limitations). This review follows that interpretation. RAE is a useful diagnostic: under a fixed protocol, were outcomes in the triggered segment better or worse than in the paired control? It is not a model-level measure of how well skills are used.

\paragraph{Two further questions, two different targets.} Trigger-conditioned results often raise two questions that require different designs. The first is the \emph{availability effect within the triggering stratum}: among runs or tasks that would trigger when the module is available, how does having the module change the outcome? Principal stratification defines this target~\cite{frangakis2002}, and assumptions such as principal ignorability given observed task features can identify it~\cite{dinglu2017}. Lu et al.\ extend principal-stratification analysis to continuous post-treatment variables~\cite{lu2026}. For runs rather than tasks, a stated coupling between arms is also needed. The analogy with complier effects in instrumental-variable analysis~\cite{angrist1996} is instructive but only partial, because the trigger is an event within a stochastic run rather than a stable property of the task. Even when identified, this target remains an availability effect. It still includes every path through which availability acts, such as metadata in the prompt, context occupancy, and behavior before the call. As an analytic example, suppose that listing skill descriptions in the prompt improves every task while the skill bodies themselves are useless. The availability effect among triggering runs is then positive, although loading a body has no effect.

The second target is the \emph{local invocation effect}: at a state where the agent would trigger, with the metadata and remaining budget held fixed, how does loading the skill compare with not loading it, given a stated policy for the rest of the run? This target is defined by intervening on the load decision itself (Section~\ref{sec:3.8}). The design must state which metadata remain visible in both branches, how the budget is accounted for, and which continuation policy is used. The two targets answer different questions, and neither can be inferred from the other.

\subsection{Linking Total and Trigger-Conditioned Contrasts}\label{sec:3.6}

Let $\pi=P(R_1=1)$ be the trigger rate under a given protocol, and let $D_{\mathrm{skip}}=\mathbb{E}[Y_1-Y_0\mid R_1=0]$ be the paired contrast on the complement. For $0<\pi<1$, the law of total expectation gives
\begin{equation}\label{eq:decomp}
\tau_{\mathrm{pair}}=\pi\,D_{\mathrm{pair}}+(1-\pi)\,D_{\mathrm{skip}},
\end{equation}
where $\tau_{\mathrm{pair}}$ is the paired mean over all tasks. Skill Following reports the empirical version of this identity as its Equation~(4), decomposing the overall skill-access effect into retrieval-invoked and skipped components~\cite{skillfollowing}. Restating the identity here makes three points explicit.

First, the identity holds for the paired statistics of any protocol and any coupling. It says nothing about the causal interpretation of its components. The decomposition in Equation~\eqref{eq:pair} applies to $D_{\mathrm{skip}}$ as much as to $D_{\mathrm{pair}}$, with the event $R_1=0$ in place of $R_1=1$. Second, the total and one component do not determine the other unless $\pi$ is also reported: a complete report gives $(\pi, D_{\mathrm{pair}}, D_{\mathrm{skip}})$ or, equivalently, $(\pi, \tau_{\mathrm{pair}}, D_{\mathrm{pair}})$. When $\pi=1$ the complement is empty and $D_{\mathrm{skip}}$ is undefined; when $\pi=0$ the same holds for $D_{\mathrm{pair}}$. Third, a non-zero $D_{\mathrm{skip}}$ is consistent with presence effects of the kind described in Section~\ref{sec:3.1}, but it does not demonstrate them, because it also contains selection terms. Attributing it to metadata in context requires an arm in which only the metadata is present, an ablation proposed in Skill Following~\cite{skillfollowing}. The Regression Tax study documents presence-only flips through trajectory evidence, namely regressions in which no skill body was read~\cite{regressiontax}. This is valuable mechanistic evidence, but the study runs each condition once per task, and one author assigned the mechanism labels; it does not include a metadata-only arm.

\subsection{Discordance and Degradation}\label{sec:3.7}

With binary success and one run per arm per task, every task falls into one of four cells. Let $\hat p_+$ be the share of tasks that fail without the module and succeed with it, and $\hat p_-$ the share that succeed without it and fail with it. Then $\hat p_+-\hat p_-$ equals the paired difference in success rates. This identity is exact, and reporting both terms rather than their difference is informative~\cite{regressiontax,vasudev2026}. The same paired binary table underlies McNemar's test, whose null concerns equality of marginal outcome probabilities~\cite{dror2018}. That testing target is distinct from identifying tasks with negative expected treatment effects. The difficulty is interpretation. It is tempting to read $\hat p_-$ as the share of tasks the module harms. That reading is not supported.

The marginal success rates of the two arms do not determine their joint distribution. This is related to the identification problem in estimating the fraction of individuals who benefit or are harmed from marginal potential-outcome distributions, for which Huang et al.\ derive bounds~\cite{huang2017}. Here the joint distribution additionally depends on the chosen coupling of stochastic executions. If both arms succeed with probability $0.5$ on a task, the chance that the pair shows a regression is $0$ when the runs are perfectly coupled in the same direction, $0.25$ when they are independent, and $0.5$ when they are perfectly anti-coupled. Under a protocol that runs the arms independently given the task, the expected observed regression rate is
\begin{equation}\label{eq:disc}
\mathbb{E}[\hat p_-]=\mathbb{E}_x\bigl[\mu_0(x)\,\bigl(1-\mu_1(x)\bigr)\bigr],
\end{equation}
which is positive even when $\mu_1=\mu_0$ for every task. Repeating runs estimates this protocol-specific rate more precisely, and a same-configuration noise floor, as in the Replay Gap study's control forks~\cite{replaygap}, shows how much of it arises without any treatment difference. Neither step identifies how many tasks are causally harmed.

This distinction separates two quantities. \emph{Paired discordance} is the observed frequency of gains and regressions under a stated protocol, with its coupling and number of runs. \emph{Task-level degradation share} is
\begin{equation}\label{eq:degr}
\delta_\varepsilon=P_x\bigl(\mu_1(x)-\mu_0(x)<-\varepsilon\bigr),
\end{equation}
the share of tasks in the population whose expected outcome falls by more than a margin $\varepsilon$. For a finite benchmark of $n$ evaluated tasks, the corresponding quantity is $\delta_{\varepsilon,n}=n^{-1}\sum_{i=1}^{n}\mathbf{1}\{\mu_1(x_i)-\mu_0(x_i)<-\varepsilon\}$. Under a stated task population and a model of repeated runs, $\delta_\varepsilon$ is estimable. Its estimation error, however, depends on the number of runs per task and on how many tasks have effects close to the margin, and no fixed small number of runs makes a naive estimate reliable.

\paragraph{An analytic counterexample.} Let both arms succeed with probability $0.5$ on every task, so that $\delta_0=0$. Run each arm three times per task, independently, and count the tasks whose treated sample mean is below the control sample mean. The count of successes in each arm is Binomial$(3, 0.5)$; the two counts are equal with probability $20/64$, so the treated count is lower with probability $22/64\approx 34\%$. The plug-in estimate of $\delta_0$ is therefore centered near $0.34$ rather than $0$, and adding tasks makes it more stable without removing the bias.

Two reporting options avoid this problem without requiring a stable point estimate from every paired study. One is to construct simultaneous confidence intervals $[L_x,U_x]$ for each task's difference $\mu_1(x)-\mu_0(x)$ and to report two shares: the share of tasks that are \emph{confidently degraded} ($U_x<-\varepsilon$) and the share that are \emph{possibly degraded} ($L_x<-\varepsilon$). When all intervals cover their corresponding task effects, these two shares bound $\delta_{\varepsilon,n}$ for the evaluated benchmark. Extending the bound to $\delta_\varepsilon$ for a wider task population additionally requires a stated sampling design for the tasks and a population-level interval. The other is a hierarchical model with stated assumptions about the distribution of task-level effects. Among the core studies, the Regression Tax study runs each condition once per task and states that run-to-run variance is not estimated~\cite{regressiontax}. Vasudev et al.\ report recovery and disruption counts from paired runs, with per-seed results~\cite{vasudev2026}. Both report paired discordance. No study was found in the review sample that reports $\delta_\varepsilon$ or either of the two shares.

\subsection{State-Level Intervention Value}\label{sec:3.8}

Activation gates decide, at a particular state, whether to load a candidate skill~\cite{skillapt}. The quantity they aim to approximate must be defined as an intervention, not as a difference between observed groups. Let $\varphi$ denote a fixed policy for the rest of the trajectory, including its remaining budget, and let $C$ denote total cost from the decision onward. Define
\begin{equation}\label{eq:q}
Q^{\varphi}_Y(s,a)=\mathbb{E}\bigl[Y^{a,\varphi}\mid S=s\bigr],\qquad
Q^{\varphi}_C(s,a)=\mathbb{E}\bigl[C^{a,\varphi}\mid S=s\bigr],
\end{equation}
\begin{equation}\label{eq:v}
v^{\varphi}(s,k)=\bigl[Q^{\varphi}_Y(s,\mathrm{load}\ k)-Q^{\varphi}_Y(s,\mathrm{abstain})\bigr]-\lambda\bigl[Q^{\varphi}_C(s,\mathrm{load}\ k)-Q^{\varphi}_C(s,\mathrm{abstain})\bigr],
\end{equation}
where $Y^{a,\varphi}$ and $C^{a,\varphi}$ are the outcome and cost when action $a$ is taken at state $s$ and $\varphi$ is followed afterwards. The cost term is the difference in total subsequent cost, not only the cost of loading $k$.

When $v^{\varphi}$ is estimated from logged runs, success and cost must each be identified, and the corresponding observed quantity is
\begin{equation}\label{eq:vobs}
\bigl\{\mathbb{E}[Y\mid s,\mathrm{load}\ k]-\mathbb{E}[Y\mid s,\mathrm{abstain}]\bigr\}-\lambda\bigl\{\mathbb{E}[C\mid s,\mathrm{load}\ k]-\mathbb{E}[C\mid s,\mathrm{abstain}]\bigr\}.
\end{equation}
A success difference alone corresponds to the first bracket of Equation~\eqref{eq:v}, not to $v^\varphi$; with a success difference of $0.1$ and a weighted cost difference of $0.2$, for instance, $v^\varphi=-0.1$. Equation~\eqref{eq:vobs} equals $v^\varphi$ when the logged runs in both arms follow $\varphi$ after the decision, and when three further conditions hold: consistency between the logged actions and the interventions; exchangeability of the two actions at the current decision given $s$; and positivity, meaning that both actions occur at every state of interest. If the logs continue under a different policy, the raw means in Equation~\eqref{eq:vobs} estimate the value of that logging continuation, not of $\varphi$, and they must be replaced by an off-policy identification formula and estimator for $\varphi$. Such formulas require, in addition, that the recorded history suffices for sequential exchangeability at every later decision and that the logging policy supports the actions $\varphi$ takes. Methods for single-step contextual bandits~\cite{li2011,dudik2011} illustrate the idea of correcting for the behavior distribution but do not by themselves cover these multi-step conditions. Jiang and Li extend doubly robust evaluation to sequential decisions by combining value estimates with target-to-behavior action-probability ratios across successive steps~\cite{jiang2016}. Applying this estimator class to agent logs still requires the stated history, support, and continuation-policy conditions. Exchangeability at the current decision alone does not remove differences in how the rest of the run unfolded. Randomizing the action at the decision point secures exchangeability by design, but the continuation condition still applies. Branching both actions from a recorded state and continuing each in closed loop also secures exchangeability, provided that the full state can be restored, the environment is reset validly, and the continuation policy is stated.

Among the core studies, SkillApt is the closest approximation of this quantity for skill activation. It defines potential outcomes for an execution state and builds evidence from matched WITH and WITHOUT executions that share task, model, decoding, environment, and evaluator~\cite{skillapt}. In its frozen confirmatory evaluation, the state is represented by hashed features of the task question, so the learned utility is conditional on the task rather than on a mid-trajectory state. Correctness is primary and costs serve only as tie-breaks. Pairs in which either arm suffers an infrastructure failure are censored, and two persistent no-skill timeouts left 111 of 113 confirmatory states complete, so the confirmatory result is a complete-case analysis. Confidence-gated retrieval with matched trajectory replay illustrates the difference between this quantity and neighboring ones~\cite{confgated}. It holds candidate answer states, evidence points, budgets, and costs fixed and compares confidence-to-action controllers on these fixed paths, which is a controller swap. Calibrating the probability that the current answer is correct, comparing controllers on fixed paths, and estimating the incremental value of another retrieval are three different targets. The study addresses the first two and concludes that the third requires a separate value-of-information estimate. It therefore shows why calibration cannot substitute for estimating $v^\varphi$.

A different decision is \emph{execution selection}: whether to run the downstream agent at all on a given query and retrieved bundle. RADEG is of this kind~\cite{radeg}. Its score predicts the probability that executing the agent on the query--bundle pair yields a non-zero verifier reward, and its gate decides whether to launch that execution (its Section~4.1). The gate does not choose between loading and not loading a skill within one run, so it does not estimate $v^\varphi(s,k)$: there is no baseline in which the same state continues without the skill. If a skipped execution is assigned zero reward, the gate targets a well-defined execution-selection utility, and the probability of non-zero reward, the expected reward, and the cost of execution must then be kept apart. The bundle-perturbation study that motivates RADEG is a separate treatment contrast, a comparison of bundle compositions on the same query, and should not be conflated with the gate's prediction target. Because the gate does not alter the executed agent's internal policy, its policy value can be evaluated from logged executions when the logs support both decisions, the execution mechanism is fixed, tasks are independent, and the reward feedback follows the stated protocol. RADEG's evaluation on logged rollouts with held-out splits defined at the query level is of this type. It differs in kind from replaying trajectories in which an intermediate action has been changed (Section~\ref{sec:4.5}).

For all such gates, a state- or query-level estimate must be followed by an evaluation of the total effect, or the policy value, of the policy the gate induces, because accurate local estimates do not guarantee a beneficial policy~\cite{vasudev2026}.

\section{Evidence Synthesis}\label{sec:4}

Five initial published works establish concrete examples of retrieval-to-outcome comparisons, memory and action-space changes, resource trade-offs, and protocol stability (Table~\ref{tab:published}). Thirteen frontier cases extend this analysis to trigger conditioning, paired discordance, gating, and replay (Table~\ref{tab:matrix}). Seventeen further detailed studies strengthen the comparisons of retrieval policies, module removal, memory horizons, execution costs, and reliability or risk protocols. Their publication labels do not rank methodological quality. Appendix Table~\ref{tab:appendix} records all thirty-five focal designs, while Appendix Table~\ref{tab:coverage_sources} retains the thirty-five targeted readings.

\begin{table*}[!htbp]
\centering\footnotesize
\caption{Selected published comparisons and their interpretive boundaries}\label{tab:published}
\begin{tabularx}{\linewidth}{>{\raggedright\arraybackslash}p{2.6cm}>{\raggedright\arraybackslash}p{4.1cm}>{\raggedright\arraybackslash}X}
\toprule
\textbf{Study / venue} & \textbf{Comparison} & \textbf{Interpretive boundary} \\
\midrule
ToolRet~\cite{toolret}; Findings ACL 2025 & Retrieved vs.\ oracle toolsets; trained vs.\ untrained retrievers with downstream agents & Supports tested replacements, not a general mapping from recall to success \\
AWM~\cite{awm}; ICML 2025 & Workflow memory vs.\ baseline agents; offline and online variants & Adaptation package and task sequence matter; observed token overhead is not a common resource cap \\
ASI~\cite{asi}; COLM 2025 & Static agent, text-skill AWM, and programmatic skills; verification/representation ablations & Main gain bundles induction, verification, and action-space changes; one high-level step may contain several primitive actions \\
AI Agents That Matter~\cite{agentsmatter}; TMLR 2025 & Agent architectures vs.\ retry baselines; cost--accuracy trade-offs & Resource comparison under stated tasks and prices, not a skill-specific invocation effect \\
StableTool\-Bench \cite{stabletoolbench}; Findings ACL 2024 & API and evaluator changes; a fixed solvable-task subset & Changes the measurement environment and population; repeated grading is not repeated execution \\
\bottomrule
\end{tabularx}
\end{table*}

\begin{table*}[!htbp]
\centering\footnotesize
\caption{Frontier case studies (A: verified acceptance; P: preprint): what each design compares and the condition that most limits its interpretation. Appendix Table~\ref{tab:appendix} gives arms, populations, runs, censoring, endpoints, versions, and locations in each paper.}\label{tab:matrix}
\begin{tabularx}{\linewidth}{>{\raggedright\arraybackslash}p{2.4cm}>{\raggedright\arraybackslash}p{3.7cm}>{\raggedright\arraybackslash}X}
\toprule
\textbf{Study} & \textbf{Quantity (review interpretation)} & \textbf{Condition that limits interpretation} \\
\midrule
Skill Following~\cite{skillfollowing} [A] & Overall paired effect; trigger-conditioned paired contrast (RAE) & Restricts to tasks where the enabled run returned a skill; shared-seed coupling with unknown selection terms \\
Regression Tax~\cite{regressiontax} [P] & Total availability effect; paired discordance & One run per task and condition; variance not estimated \\
Budget study~\cite{worththeirtokens} [A] & Comparison under an approximate budget constraint & Budget matched through a step cap, not exactly; control arm also adds pruning \\
Skill shadowing~\cite{shadowing} [P] & Composition effect with counterfactual decomposition & Population restricted to task--model pairs whose skills each raised pass rate by at least 4 percentage points; bounds need a monotonicity assumption \\
SkillsBench \cite{skillsbench} [A] & Total availability effect of curated per-task bundles & Task-specific bundles supplied in the evaluation; not retrieval from a shared library \\
SkillApt~\cite{skillapt} [P] & Task-conditional load vs.\ abstain utility & State represented by question features; complete-case analysis (111 of 113 states) \\
RADEG~\cite{radeg} [P] & Execution selection: predicted probability of non-zero reward & Gate decides whether to run the agent, not whether to load a skill; each pair executed once \\
Confidence-gated retrieval~\cite{confgated} [P] & Comparison of confidence-to-action policies & Fixed evidence paths; does not estimate the value of another retrieval \\
Replay Gap~\cite{replaygap} [A] & Validity of replay for per-step model switching & Tested for model switching only \\
Failure prevention~\cite{vasudev2026} [P] & Total effect of intervention; recovery and disruption counts & Counts are paired discordance, not shares harmed \\
State-Path menu~\cite{menu} [A] & Component-swap total effect & One executor with deterministic decoding for the main result \\
Capability Pages~\cite{capabilitypages} [P] & Component-swap total effect & Retrieval and use contributions not separated \\
TabAgent~\cite{tabagent} [P] & Component-swap total effect with cost & One benchmark and decision head for the task-level result \\
\bottomrule
\end{tabularx}
\end{table*}

\subsection{Evaluation Settings and Target Populations}\label{sec:4.settings}

The benchmark defines the scope of an agent comparison. Mind2Web evaluates action predictions on recorded web states with ground-truth action history and separates generalization across tasks, websites, and domains~\cite{mind2web}. Its whole-task success requires every independently evaluated step to be correct; it is not a new closed-loop execution. WebArena instead provides executable websites and checks task completion from the resulting states and outputs~\cite{webarena}. The distinction matters for AWM, which is evaluated on both benchmarks~\cite{awm}: an improvement on recorded-state action prediction and an improvement on interactive task completion are complementary findings, not two estimates of an identical endpoint.

Task scope also changes within executable environments. WorkArena samples instances of knowledge-work tasks on ServiceNow and supplies task validation and oracle routines~\cite{workarena}. OSWorld covers desktop and web applications with task-specific initial states and execution-based checks~\cite{osworld}. AgentBench combines several interactive environments under a common agent-evaluation framework~\cite{agentbench}. These resources broaden the settings in which a skill system can be tested, but benchmark breadth does not identify a component effect. That still requires specifying which component changes within each environment. Likewise, a result on one workflow family does not establish transport to another without a stated target population and evidence about the changed interfaces and tasks.

Tool-use benchmarks differ in the information supplied and in what counts as success. API-Bank separates calling, retrieval plus calling, and planning plus retrieval plus calling~\cite{apibank}. BFCL uses different checks for its categories, including syntax-tree matching for function calls and combined state and response checks for multi-turn tasks~\cite{bfcl}. ToolSandbox evaluates interactive trajectories against required milestones and forbidden events, while $\tau$-bench evaluates database outcomes and required responses after interaction with a simulated user~\cite{toolsandbox,taubench}. Function-call, trajectory-milestone, and task-completion scores therefore require distinct interpretations. Even a task reward may not capture every relevant constraint: the $\tau$-bench authors explicitly note that the correct final outcome can coexist with a policy violation, such as acting without required confirmation. These distinctions clarify the population and outcome axes of Table~\ref{tab:axes}.

Benchmark scores can summarize different degrees of completion. WebShop distinguishes a graded reward for satisfying product constraints from success on the full request, while ScienceWorld gives credit for task-specific subgoals~\cite{webshop,scienceworld}. ALFWorld connects text-based tasks with embodied execution, making the observation and action interface part of the setting~\cite{alfworld}. AgentGym combines environments with their own success or reward measures~\cite{agentgym}. Consequently, an aggregate across such environments depends on the task mixture and score definitions; it cannot be interpreted as a common probability of successful skill use without aligning those definitions.

Interface and information changes also alter the comparison. VisualWebArena adds tasks that require visual grounding, and AndroidWorld evaluates parameterized tasks in a controlled mobile environment~\cite{visualwebarena,androidworld}. WorkArena++ contrasts explicit workflow instructions with ticket-based goals that require consulting a knowledge base~\cite{workarenapp}. SWE-agent studies an agent--computer interface for repository tasks~\cite{sweagent}. A gain after adding a reusable procedure may therefore depend on what the interface already reveals, what workflow knowledge the instructions supply, and what operations the agent can execute. These factors belong in the treatment and population descriptions.

Answer-oriented benchmarks introduce another boundary. GAIA grades final answers to questions that may combine reasoning, browsing, and tool use; AssistantBench evaluates information-seeking answers on the web and separately examines abstention~\cite{gaia,assistantbench}. ToolQA constructs questions from external data sources with accompanying tools~\cite{toolqa}. These designs test whether an agent reaches an accepted answer under the specified resources. Answer correctness does not alone establish that a particular retrieved tool or skill caused the success, and precision among answered questions differs from performance over all assigned questions.

\subsection{Exposure and Candidate Construction}\label{sec:4.1}

Candidate construction defines the information and actions available downstream. ToolRet evaluates both item relevance and Completeness@k, which asks whether all labeled target tools occur in the shortlist~\cite{toolret}. This is a published example of matching the component metric to tasks requiring multiple tools. Its merged corpus also raises the possibility that tools outside the original labels can solve a query, a limitation discussed by the authors. Coverage of a reference set therefore differs from coverage of all valid solutions.

Re-Invoke evaluates document expansion and query-intent rewriting for single- and multi-tool retrieval, and separately compares downstream tool use on six ToolBench subsets with ToolLLaMA and DFSDT held fixed~\cite{reinvoke}. The reported pass rates, reproduced using GPT-3.5-turbo grading, compare Re-Invoke, the trained ToolLLM retriever, and supplied reference tools without retrieval. That last arm still provides tools. The downstream comparison supports retriever replacement under the specified executor and evaluation protocol; it neither supplies a no-tools contrast nor makes ranking gains a generally valid surrogate for execution success. Shortlist length should likewise be treated as a design choice: changing it changes both candidate coverage and the context the agent receives. A relevance score alone cannot decide that trade-off or measure unsafe exposure.

Tool selection is not always an external retrieval operation. ToolkenGPT learns embeddings that let a frozen language model select tools during token generation~\cite{toolkengpt}. Its intervention includes learned calling behavior; it differs from replacing the retriever while keeping the executor fixed. CRAFT creates and verifies a specialized toolset before retrieving tools for inference~\cite{craft}. CREATOR separates tool creation, decisions about use, execution, and rectification~\cite{creator}. In these systems, tool quality and construction rules are part of the treatment. Comparing the full system with a baseline informs deployment of that package; attributing the gain specifically to retrieval additionally requires comparisons that hold the generated tools and execution procedure fixed.

ToolGen represents tools as vocabulary items and trains a model to retrieve and invoke them through generation~\cite{toolgen}. ToolACE generates and verifies tool-use training dialogues, changing the data from which calling behavior is learned~\cite{toolace}. These are interventions on the representation or training of a tool-using policy. Their evaluation contrasts differ from an inference-time retriever swap with an unchanged executor; training data, model updates, and candidate representation should be recorded as parts of the intervention.

Planning systems can change several decisions together. Chameleon assembles tools into a program for a task, whereas HuggingGPT separates planning, model selection, execution, and response generation~\cite{chameleon,hugginggpt}. AnyTool combines hierarchical API selection with self-reflection~\cite{anytool}. Agent Lumos separates planning, grounding, and execution, and ToolPlanner combines candidate-tag extraction and path planning with task-completion and instruction-following feedback~\cite{agentlumos,toolplanner}. These systems provide distinct ways to organize a tool-using policy. A comparison of each complete system with its baseline estimates the performance of that package; attribution to selection alone requires keeping the planner, training procedure, execution, and recovery rules aligned.

The frontier State-Path menu study tests a more specific hypothesis: constructing menus around executable routes improves downstream success~\cite{menu}. Its comparison with an unchanged executor supports that menu replacement in the tested setting. It does not establish that chain coverage is a universally sufficient surrogate for task success. This distinction between a suitable component metric and a validated outcome surrogate also applies to skill retrieval.

\subsection{From Retrieval Gains to Outcome Gains}\label{sec:4.2}

ToolRet provides a published retrieval-to-outcome comparison: its ToolBench experiments replace oracle toolsets with retrieved ones and compare retrievers before and after training, with GPT-3.5 and ToolLlama as executors~\cite{toolret}. The results connect better retrieval with better outcomes in those configurations. They do not isolate the causal path through ranking quality from other changes in the supplied toolset. The frontier Capability Pages study gives a narrower representation contrast: including negative-boundary text improves Recall@10 across its tested retrievers and raises mean end-to-end success by 3.62 percentage points with shared executors~\cite{capabilitypages}. Both designs inform component choice within a package; neither supplies a no-module comparison by itself.

Iterative tool retrieval makes the feedback policy part of selection: Xu et al.\ use language-model feedback to refine the user instruction before another retrieval step, and evaluate both retrieval and downstream tool use~\cite{iterativetoolretrieval}. Their downstream comparison uses ToolLLaMA on intra-category multi-tool instructions, whereas their broader retrieval tests cover additional distributions. The default procedure retrieves ten candidates and allows three feedback iterations. A comparison of this procedure with a single retrieval call changes the number and content of opportunities to select tools. To attribute an outcome difference specifically to candidate quality, the feedback and execution budgets must therefore be described alongside the retrieval scores.

\paragraph{Curated provision and retriever replacement.} A component-swap effect does not show that skills help relative to no skills. A better retriever can raise success over a weaker one while the whole module still lowers success relative to no module. SkillsBench instead compares no skills with curated, task-specific bundles, which raise the macro pass rate from 33.9\% to 50.5\% across 18 configurations~\cite{skillsbench}. These bundles are supplied for the evaluated tasks, so the result concerns their availability, not retrieval from a shared library. The same study reports that skills the agent generates for itself fall below the no-skills baseline on the three configurations where this condition was run. Availability effects therefore depend strongly on what the library contains.

Read together, SkillsBench and the retrieval studies support distinct decisions. Curated provision tests whether a particular bundle helps when supplied; retrieval from a shared library tests the deployed selection package against its stated baseline; replacing a retriever or its representations tests which component performs better within that package. Capability Pages, for example, compares representations with and without negative-boundary text while holding the executor fixed~\cite{capabilitypages}. Its gain supports that replacement in the evaluated setting. It neither estimates the no-skills contrast in SkillsBench nor transports SkillsBench's gain to shared-library retrieval. A study addressing both deployment and component choice would need a no-skills arm as well as the alternative retrieval arms on the same task population. The present cross-study evidence motivates those contrasts but cannot supply their missing outcomes.

\paragraph{Attribution and metric transfer.} An observed association between recall gains and outcome gains does not identify the path that produced the improvement. An outcome gain that accompanies a recall gain is consistent with better candidates being used, but also with changes in context length, formatting, or the presence of different text~\cite{regressiontax}. The authors of Capability Pages explicitly present their product-form success model as an attribution sketch rather than an identity~\cite{capabilitypages}. The published ASI ablations separate aspects of verification, representation, and placement in memory or the action space on the evaluated shopping website~\cite{asi}. They support those local contrasts, although the main ASI-versus-AWM comparison changes several elements together. The invocation-conditioned decomposition of Song and Wei~\cite{shadowing} and the trigger-conditioned pairing of Skill Following~\cite{skillfollowing} offer further diagnostics, subject to the qualifications of Sections~\ref{sec:3.5} and~\ref{sec:3.6}.

Other comparisons show that an improved local metric need not yield a proportionate workflow improvement. In a customer-support study, gold-history next-turn evaluation showed improved all-turn success for all four fine-tuned models, while deterministic replay completion reached at most 8/77 tool-requiring workflows (10.4\%) and holistic success was 0/77 in every configuration~\cite{betterturns}. The replay generated assistant histories but used fixed gold user turns and exact reference tool-call matching. These constraints can reject valid alternatives and limit interpretation as live-user deployment success. TabAgent illustrates a further outcome profile: it reports maintained task-level success on AppWorld alongside lower shortlisting latency and inference cost~\cite{tabagent,appworld}.

\subsection{Activation and Actual Use}\label{sec:4.3}

A surfaced candidate is not necessarily loaded, and a loaded candidate is not necessarily used well. Recent work separates these levels.

\emph{Retrieval versus activation.} SkillApt treats ``which skill is relevant'' and ``whether loading it is worthwhile'' as separate decisions. On its frozen SRA-Bench evaluation it reports the same observed accuracy as BM25 Top-1 (0.838 vs.\ 0.838) while cutting the activation rate from 100\% to 31.5\%~\cite{skillapt}. SkillApt targets a task-conditional value of loading (Section~\ref{sec:3.8}). For confidence-gated retrieval in question answering, matched trajectory replay compares confidence-to-action controllers on fixed paths and shows that calibration changes which questions are answered without estimating the benefit of another retrieval~\cite{confgated}; it is a useful counterexample to treating calibrated confidence as a value estimate. RADEG addresses a related decision: whether to run the downstream agent at all on a query and its retrieved bundle. Its gate predicts the probability of a non-zero verifier reward. The motivating study separately perturbs bundles for the same query and shows that reward is sensitive to bundle composition, while relevance scores predict it poorly~\cite{radeg}. Their shared lesson is that the value of a candidate depends on context and must be estimated from contrasts rather than inferred from relevance scores.

\emph{Memory guidance and executable actions.} AWM primarily supplies induced workflows as context, whereas ASI can execute induced skill programs~\cite{awm,asi}. ASI's verification procedure checks completion, actual use of a new skill, and changes to the environment before admitting skills. This supplies a concrete operational definition of use. It also makes clear that being available, being called, and contributing to success are separate events. A valid skill can still be unnecessary at a particular state. Comparing whole adaptation systems does not identify that state's loading or execution effect.

Chameleon's module-disabling analysis illustrates a narrower intervention than observed call frequency: it measures performance after removing selected modules from generated programs with ChatGPT on 500 test examples~\cite{chameleon}. HuggingGPT separately evaluates passing and rationality at intermediate stages and final request resolution on 130 constructed requests~\cite{hugginggpt}. Its three human raters provide repeated judgments, not three independent agent executions. ToolPlanner distinguishes matching the tool labels requested by an instruction from task completion~\cite{toolplanner}. These distinctions separate observed use, conformity to a reference procedure, and outcome change under an intervention; none should substitute for the others.

The execution interface can change what a selected capability allows the agent to do. CodeAct represents actions as executable Python code, permitting tool calls to be composed within a program~\cite{codeact}. Its M3ToolEval comparison stops after a correct answer or ten interaction turns on 82 authored instances. An action counted at that interface need not equal one atomic API call, so the turn cap does not equalize the number of tool executions. Likewise, Agentless organizes repository repair into localization, repair, and validation rather than an open-ended interaction loop~\cite{agentless}. Comparisons across these designs concern the action representation and control procedure as well as access to tools. Agentless also takes agent-baseline results from leaderboards or prior reports, rather than rerunning every baseline within one controlled harness~\cite{agentless}. Its forty candidate patches per issue are internal search samples, not forty independent evaluations of the complete system. Sharing a task set does not mean that these comparisons isolate skill activation.

\paragraph{Triggered subsets, paired flips, and expected task effects.} Skill Following reports, across 17 LLMs on coding and mathematics tasks, that models can have a positive retrieved-versus-skipped lift and a negative paired contrast on the tasks where retrieval occurred~\cite{skillfollowing}. The authors already delimit the paired contrast as protocol-conditional. The decomposition explains why: the contrast depends both on which tasks enter the subset and on which stochastic outcomes are selected within each task. The latter contribution depends on the coupling. The sign of this contrast cannot be substituted for the all-task availability effect or the effect of loading a skill at a fixed state.

Regression Tax and Failure prevention retain paired outcome changes as well as aggregate success~\cite{regressiontax,vasudev2026}. They show why an average can conceal gains on some observed pairs and losses on others. Their treatments and populations differ: one changes skill libraries in document and spreadsheet tasks, while the other adds critic interventions across question-answering and interactive tasks. Their repetition protocols also differ: Regression Tax runs each condition once per task, whereas Failure prevention uses two or three seeds depending on the configuration. These are complementary demonstrations of paired discordance, not directly comparable estimates of a common harm rate. More seeds can characterize variation in the reported counts, but the counts alone do not estimate the share of tasks with lower expected success. That target requires the per-task uncertainty treatment in Section~\ref{sec:3.7}.

Together, the three studies distinguish a triggered-segment diagnostic, an all-task outcome difference, and a distribution of task-level expected effects. A report can legitimately contain all three only if its design and uncertainty analysis support each. Regression Tax's trajectory labels also suggest presence, grounding, and verification mechanisms, but the one-run design and absence of a metadata-only arm limit causal attribution to those paths~\cite{regressiontax}.

\subsection{Library Scale, Evolution, and Cost}\label{sec:4.4}

Effects of individual skills do not add up to the effect of a library. The frontier shadowing study selects SkillsBench task--model pairs in which every authored skill raised pass rate by at least 4 percentage points over no skills (38 pairs from two models). Expanding these helpful sets to 202 skills lowers pooled pass rate by 21 percentage points~\cite{shadowing}. The version examined does not state whether selection and estimation used separate runs. The decomposition fixes either invocation probabilities or conditional pass rates, and its bounds assume that conditional pass rates under the full library do not exceed those under the helpful set. At 202 skills, the reported selection term is 0.14 (95\% interval 0.06 to 0.26), while the context term is 0.07 (interval $-0.13$ to 0.25). This is a conditional case study, not an estimate of the typical effect of scaling a skill library.

The time at which experience is acquired separates several forms of memory evaluation. Reflexion uses feedback and reflective memory to improve subsequent attempts at a task~\cite{reflexion}. ExpeL gathers experiences from training tasks, then retrieves successful example trajectories and supplies the full list of extracted insights when attempting unseen evaluation tasks~\cite{expel}. Voyager builds a library of executable programs during open-ended exploration and tests reuse in a new Minecraft world~\cite{voyager}. These designs address within-task adaptation, cross-task transfer, and continuing skill acquisition, respectively. Their repetition units also differ: the focal Reflexion experiment follows twelve adaptive attempts on ALFWorld tasks; ExpeL reports uncertainty across four validation folds; Voyager reports three trials with distinct exploration and transfer horizons~\cite{reflexion,expel,voyager}. Their results cannot be ranked as the effect of a common ``memory'' treatment without aligning the learning opportunities, evaluation population, and resource accounting. The distinction follows from their protocols; it does not invalidate the questions those protocols were designed to answer.

Other memory designs broaden the meaning of experience reuse. Self-Refine iterates feedback and revision on a current output, while A-Mem constructs linked notes that can evolve as new memories arrive and evaluates their use in long-conversation question answering~\cite{selfrefine,amem}. Generative Agents combines stored observations, reflection, and planning in a social simulation; its controlled memory ablations assess human-rated believability of interview responses under a common simulated history~\cite{generativeagents}. The one hundred evaluators rank interview responses; the ablated systems are not rerun to generate their own histories. This contrast isolates access to the existing history at interview time, rather than the end-to-end effect of changing the memory architecture during the simulation. These studies concern different memory contents, update rules, and outcomes. They motivate reporting when information enters memory and what later decisions can access it, rather than treating every improvement from retained text as transfer of an executable skill.

Online AWM and ASI likewise add skills over a sequence of tasks~\cite{awm,asi}. Their treatment is therefore an adaptation procedure, not just a final library snapshot. Task ordering, admission decisions, and previously acquired skills belong to the protocol being evaluated. A snapshot comparison and a comparison of learning procedures answer different deployment questions.

Search and scheduling introduce further resource dimensions. Tree of Thoughts searches over intermediate reasoning states, and Language Agent Tree Search combines search with environmental feedback while assuming that earlier environment states can be restored~\cite{tot,lats}. LLMCompiler schedules tool calls according to their dependencies and permits parallel execution~\cite{llmcompiler}. Its setup reports mean accuracy over three runs, and its HotpotQA and movie-recommendation latency comparisons use a ReAct baseline prompted to reduce repeated calls and early stopping; ParallelQA uses ReAct. The observed profile therefore depends on the specified control as well as the scheduling policy. Search expansions, tool invocations, generated tokens, and elapsed time measure different resources. In particular, lower latency through concurrency need not imply fewer calls, and a search-based gain should be interpreted relative to its branching and evaluation budget.

\paragraph{Total outcomes, observed cost, and budget constraints.} AI Agents That Matter compares agent architectures with simple retry strategies and reports cost alongside accuracy~\cite{agentsmatter}. That published analysis supplies the resource-comparison rationale independently of recent skill preprints. For skill systems, AWM reports a token-cost breakdown, while ASI reports action counts whose units change when a program call replaces several primitive actions~\cite{awm,asi}. Fewer such steps cannot alone establish lower total compute. TabAgent reports maintained AppWorld success with lower shortlisting latency and inference cost~\cite{tabagent}. These are observed cost--success profiles; an equal-budget claim additionally needs a common resource rule.

Construction costs introduce a further deployment choice. LATM separates an expensive tool-making stage from subsequent tool use and explicitly amortizes construction across instances of a task~\cite{latm}. This differs from comparing only the marginal cost of calling an already constructed tool. CRAFT's offline construction and ExpeL's experience collection create the same accounting question even when the learned artifacts differ~\cite{craft,expel}. A cost claim should state whether preparation is included and over how many future uses it is spread. A comparison of mature libraries need not establish the cost of deploying a new library for a short workload.

Hajimiri et al.\ instead strengthen the vanilla web-agent control by extending its actor horizon from 10 to 15 steps and adding accessibility-tree pruning~\cite{worththeirtokens}. Under this approximate resource comparison, the control matches or surpasses three online augmentation methods in aggregate success across four WebArena domains and three models, often with fewer total tokens. The authors explicitly note that the step cap does not exactly match token budgets. Because both the horizon and pruning change, this comparison assesses the augmented systems against that specified control package; it does not isolate the effect of extra steps alone. The same study separately ablates horizon extension and pruning under Gemini 3 Flash across the four WebArena domains (its Section~5.3, Table~5), reporting that the longer horizon mainly improves success while pruning mainly reduces token use~\cite{worththeirtokens}.

These comparisons do not negate gains from deploying a module under its original configuration. They distinguish an adaptation package's total outcome, its observed cost, and performance after resource reallocation. The original AWM and ASI studies and the later budget study use different agents and controls; their numerical differences cannot be subtracted to estimate the causal contribution of memory or extra steps. A matched comparison must specify the actor, environment, allocation rule, and budget in every arm.

\subsection{Validity of Evaluation Protocols}\label{sec:4.5}

The designs above depend on the protocols used to generate their counterfactual outcomes. Published protocol evidence and frontier case studies expose different sources of uncertainty.

\emph{Environment and evaluator stability.} StableToolBench changes both the API environment, using cached or simulated responses, and the evaluation procedure~\cite{stabletoolbench}. It retains 765 tasks judged solvable from 1,100 original tasks. Its main table runs each model once and grades outputs three times. Those repeated judgments characterize grading variation, not run-to-run execution variation. The resulting score concerns a selected population in a stabilized environment; it cannot be read as live-API deployment success without further validation.

\emph{Replay is not intervention.} Offline replay evaluates a changed component by substituting its output into logged trajectories while assuming that the rest of the trajectory is unaffected. Gonuguntla tested this assumption for per-step model switching on SWE-bench~\cite{swebench} with branching rollouts and same-model control forks. Swaps rewrote 61--94\% of post-fork actions, only 3\% of replayed states remained valid, and replay mispredicted every success-relevant outcome~\cite{replaygap}. The study concerns model switching, but the same logic applies to any component that changes what the agent does next: its counterfactual outcomes should come from executed continuations unless replay has been validated for that component.

\emph{Local accuracy is not intervention benefit.} A critic with an offline AUROC of 0.94 caused a 26-percentage-point collapse on one model and almost no change on another. The authors explain this with a disruption--recovery trade-off and propose a 50-task pilot to decide whether to intervene~\cite{vasudev2026}. For activation gates, a gate's offline accuracy does not determine the total effect of the policy it induces.

\emph{Stochastic execution.} In the Replay Gap study, temperature-0 serving diverged on over 90\% of control forks under one quantization configuration~\cite{replaygap}, and Hajimiri et al.\ report that run-to-run variance changed their conclusions~\cite{worththeirtokens}. Related reinforcement-learning work shows that random seeds and implementation choices can materially change reported performance~\cite{henderson2018}. Agarwal et al.\ recommend interval estimates for aggregate benchmark performance, including stratified bootstrap resampling of runs within tasks~\cite{agarwal2021}. This supports reporting aggregate uncertainty; it does not supply simultaneous per-task intervals or identify a degradation share. Single-run designs cannot separate treatment differences from this variation at the task level, which is why Section~\ref{sec:3.7} separates discordance from degradation.

\emph{Reliability differs from paired harm.} The published $\tau$-bench defines pass-hat-$k$ as the probability that all $k$ independent trials succeed, averaged over tasks, and distinguishes it from pass-at-$k$, the probability that at least one succeeds~\cite{taubench}. These are different summaries of repeated execution. Neither is a contrast between two skill configurations or the share of tasks whose expected success decreases. A reliability comparison between configurations should retain the same task population and user-simulation protocol, and it should not substitute paired gain or regression counts for the repeated-success target.

\emph{Risk depends on how scenarios are generated.} ToolEmu assesses agent behavior in an LM-emulated tool environment, including an adversarial emulator that seeks failure-inducing states~\cite{toolemu}. Its curated test cases and evaluator validation support analysis of failures under that protocol. They do not supply the frequency of those failures in an ordinary deployment population. ToolSandbox's forbidden-event checks answer a different question about specified events within its authored trajectories~\cite{toolsandbox}. Safety conclusions should therefore name both the event and the scenario-selection process, rather than equate a stress-test failure rate with operational risk.

Adversarial evaluations also define distinct target populations. InjecAgent starts from a constructed tool-response state, assuming a correct initial tool call, and evaluates subsequent calls for following an embedded attacker instruction~\cite{injecagent}. It reports both ASR-valid, which excludes invalid outputs from the denominator, and ASR-all, which includes them; these answer different questions about the evaluated agents. AgentDojo evaluates attacks and defenses within stateful tasks, alongside utility on the user's task~\cite{agentdojo}. Its metric definitions distinguish benign user-task utility from attack outcomes on user-task/injection-target pairs. The actual aggregation weights must therefore be checked before interpreting a difference as a paired utility effect. AgentHarm separately scores progress on malicious multi-step tasks and refusal using synthetic tools without real side effects~\cite{agentharm}. These protocols distinguish resistance to external instructions, preservation of legitimate task utility, and willingness to execute harmful workflows. The resulting rates depend on the attack distribution, injection opportunity, and scoring rules; they do not estimate the prevalence of harm in ordinary deployment.

\emph{Off-policy estimation needs support.} Classical work on replay evaluation and doubly robust policy evaluation establishes conditions for using logged outcomes~\cite{li2011,dudik2011}. Actions required by the target policy need support under the logging policy, along with the relevant consistency and propensity assumptions. For example, the consistency guarantee studied by Thomas and Brunskill assumes that an action with zero probability under a behavior policy also has zero probability under the evaluation policy, alongside bounded-weight and other conditions~\cite{thomas2016}. Applied to a skill gate, this requires stating whether logs cover both loading and abstaining and whether subsequent behavior remains governed by the specified continuation policy. An accurately predicted outcome under the logged action is not automatically an identified value for an unsupported alternative.

Finally, outcome definitions should reflect the intended task. AgentBoard adds progress measures based on state matching or annotated subgoals alongside final success~\cite{agentboard}. ASI's scaled-up activity experiments use intermediate checkpoints, whereas its ordinary WebArena evaluation uses task success~\cite{asi}. Such progress measures can distinguish partially successful trajectories without establishing that the entire task was completed. Checkpoint completion, reference-tool coverage, repeated-success reliability, and final success are legitimate but different endpoints. The recommendation is to state which endpoint the decision requires rather than interpret all of them as the same measure of agent capability.

\section{Discussion}\label{sec:5}

\subsection{Cross-cutting Lessons}\label{sec:5.1}

\textbf{Name the treatment and the question.} In current papers, ``skills'' can mean making metadata visible, retrieving with a particular retriever, loading a skill, or following it, and a comparison can ask whether to deploy a module, which component to prefer, or what to do at a given state. These are different contrasts with different answers. A paper should state which contrast it manipulates and which question it answers before reporting results.

\textbf{Treat post-treatment strata with care.} Pairing on the task removes between-task selection but not within-task selection (Section~\ref{sec:3.5}). A trigger-conditioned contrast should be reported together with the trigger event, the trigger rate, and the complementary contrast, and it should be interpreted as a protocol-specific diagnostic unless the coupling between arms and the identifying assumptions are stated. Even when identified, an effect within the triggering stratum is an availability effect; a claim about invoking a skill needs a design that intervenes on the load decision.

\textbf{Report discordance as discordance.} Counts of gains and regressions are informative~\cite{regressiontax,vasudev2026}, but they describe a protocol, not the share of tasks harmed (Section~\ref{sec:3.7}). Claims about harm to tasks need repeated runs per arm and interval-based summaries, such as the shares of confidently and possibly degraded tasks, or a stated model for task-level effects.

\textbf{Distinguish total effects from budget-constrained comparisons.} Both are legitimate. A total effect answers whether to deploy a module as specified; a budget-constrained comparison answers which configuration to choose under a cap. Resource-efficiency claims need the latter, or a cost--success frontier~\cite{agentsmatter,worththeirtokens}.

\textbf{Match component metrics to task needs.} ToolRet's completeness measure illustrates set coverage, while its downstream experiment tests the resulting agent separately~\cite{toolret}. Coverage of labeled tools, coverage of valid solutions, and safe exposure should not be treated as interchangeable.

\textbf{Execute counterfactuals, or validate replay.} Fixed-trajectory replay, gold-history scoring, and model-history execution against fixed reference user turns impose different constraints~\cite{replaygap,betterturns}. Their validity for a deployment claim depends on which actions and responses can change after intervention. When full closed-loop evaluation is too costly, a small closed-loop pilot can at least check the sign of an intervention's effect~\cite{vasudev2026}.

\subsection{Reporting Checklist}\label{sec:5.2}

Table~\ref{tab:checklist} lists the information recommended in this review for papers that claim a tool or skill component improves an agent. Each item states the designs to which it applies. Most items can be reported from runs that authors already perform.

\begin{table}[!htbp]
\centering\small
\caption{Reporting checklist for tool and skill evaluation}\label{tab:checklist}
\begin{tabularx}{\linewidth}{>{\raggedright\arraybackslash}p{2.6cm}>{\raggedright\arraybackslash}X>{\raggedright\arraybackslash}p{3.3cm}}
\toprule
\textbf{Item} & \textbf{What to report} & \textbf{Applies to} \\
\midrule
Question and contrast & Deployment, component choice, or state-level decision; what each arm sees & All designs \\
Target population & Task distribution and whether it matches the query distribution of any retrieval benchmark used & All designs \\
Unit, coupling, runs & Unit of comparison; how arms are coupled (independent, shared seed, branched state); runs per task and arm; run-to-run variance & All designs with stochastic agents \\
Identification & Assumptions under which the reported summary is causal, or a statement that it is protocol-specific & Designs that condition on post-treatment events or use logs \\
Trigger stratification & Trigger rate $\pi$ and paired contrasts on triggered and non-triggered tasks, with the trigger definition & Designs in which retrieval or invocation is optional \\
Discordance and degradation & Gain and regression counts under the stated protocol; if harm to tasks is claimed, confidently and possibly degraded shares from simultaneous per-task intervals, or a stated hierarchical model & Paired designs; degradation shares only for harm claims \\
Budget & Tokens, calls, and latency per arm; for efficiency claims, an ex-ante cap with allocation rule, or a cost--success frontier & All designs; cap only for efficiency claims \\
Protocol & Closed loop, branching, or replay; if replay, evidence that it is valid for the component & Designs using logged trajectories \\
Component metrics & Set or prerequisite coverage and risk exposure, in addition to single-item recall & Retrieval and menu studies \\
Library state and preparation & Library size, version, admission rules, and prior task exposure; construction cost and amortization horizon when efficiency is claimed & Studies of learned or changing libraries \\
\bottomrule
\end{tabularx}
\end{table}

\paragraph{Applying the checklist.} The following examples use the source locations recorded in Appendix Table~\ref{tab:appendix}. They illustrate interpretation, rather than score the studies or validate the checklist.

\emph{Coupling and identification.} Skill Following's Section~4.3 reports shared task prompts, generation seeds, and decoding configurations; its Limitations explicitly delimit RAE as protocol-conditional~\cite{skillfollowing}. This supports a diagnostic for the triggered subset under that pairing protocol. A local invocation claim would additionally need an intervention on loading at a fixed state, with metadata, remaining budget, and continuation policy specified. The distinction concerns the targets; it does not imply that the source authors made an unacknowledged causal claim.

\emph{Discordance and degradation.} Regression Tax's Sections~3.4--3.6 and~6.2 report one run per condition and task and state that run-to-run variance is not estimated~\cite{regressiontax}. The paired counts support a description of flips in those executions. A claim about the share of tasks whose expected success decreases would require repeated per-task evidence with simultaneous uncertainty bounds or a stated model; repeating the same discordance summary would not by itself support that claim.

\emph{Budget.} The web-agent budget study's Sections~3.2--3.3, 5.1, and~5.3 and Appendices~B--D specify the longer actor horizon, pruning, repeated runs, and approximate token comparison~\cite{worththeirtokens}. These support a comparison with that control package. An exact equal-budget claim would additionally need a common ex-ante resource cap and allocation rules for every arm. This reporting distinction preserves the study's own qualification of its budget control.

\subsection{Open Problems}\label{sec:5.3}

\emph{Invocation effects with run-level triggers.} The trigger is an event within a stochastic run, not a fixed property of a task. Principal-stratification methods identify availability effects within strata under assumptions such as principal ignorability~\cite{dinglu2017}. No study was found in the review sample that states a potential-outcome model for runs that identifies a stratum effect or a local invocation effect while specifying how triggering varies across runs of the same task and how the arms are coupled. Developing such models for agents and designs that intervene on the load decision at scale remains an open problem.

\emph{Affordable valid counterfactuals.} Branching rollouts and paired closed-loop runs are expensive~\cite{replaygap}. Methods that decide when replay is safe, or that allocate a small number of closed-loop runs to the most informative states, would make valid evaluation cheaper.

\emph{Benchmarks with built-in identification.} Benchmarks could include a metadata-only arm, with skills visible in the prompt but not retrievable, alongside no-skill and full-access arms. Branching from saved states under fixed continuation policies would support local loading comparisons. Such controls would make component attribution a designed comparison rather than an interpretation added after observing success rates.

\emph{Evaluating evolving libraries.} When libraries evolve during deployment~\cite{awm,asi}, the natural target is an admission policy rather than a fixed snapshot. How to evaluate an admission policy without re-running the entire evolution is an open problem.

\emph{From state-level values to policy effects.} Activation gates trained on state- or task-conditional utilities~\cite{skillapt} and execution gates trained on predicted rewards~\cite{radeg} induce policies whose total effect or policy value, and degradation share, must be evaluated separately~\cite{vasudev2026}. A link between the accuracy of the utility estimates and the degradation share of the resulting policy would connect activation research with evaluation.

\subsection{Limitations of This Review}\label{sec:5.4}

The thirty-five studies examined in detail were selected for their relevance to the methodological argument rather than to represent publication frequencies, which may leave some evaluation designs underrepresented.

Experiments were not reproduced, and replication cannot be inferred from publication.

\section{Conclusion}\label{sec:6}

Tool and skill components are currently evaluated with a mix of retrieval metrics, component-swap comparisons, retrieved-versus-skipped contrasts, trigger-conditioned paired contrasts, gain and regression counts, and budget-constrained comparisons, often under the same terminology. These designs address different questions. Some are descriptive, some estimate total effects of deploying or swapping a component, and some describe one protocol without identifying a mechanism. Three interpretations in particular need care. Pairing on the task does not turn a trigger-conditioned contrast into an invocation effect. Regression counts from independent runs measure discordance rather than harm. A comparison without budget matching estimates the total effect; the absence of matching does not make that effect invalid. Stating the contrast, population, unit and coupling, budget, summary, and identification assumptions of each design costs little, and helps readers compare results that currently appear to conflict. Agents now draw much of their capability from what they retrieve. Evaluating that retrieval therefore requires stating precisely what each design shows.

\appendix
\section{Study Evidence and Reading Register}\label{app:evidence}

This appendix separates detailed extraction of focal evaluation designs from targeted readings used to define the surrounding architectures, settings, and outcomes. The groups are disjoint: thirty-five studies appear in Table~\ref{tab:appendix} and thirty-five in Table~\ref{tab:coverage_sources}. Neither grouping is a quality ranking. Section~\ref{sec:2} describes the roles of the thirty other papers and two non-paper sources.

\subsection{Detailed Study-Level Evidence}\label{app:matrix}

Table~\ref{tab:appendix} records the source version, arms, population, repetition, budget, endpoint, and locations supporting each focal comparison. P denotes a preprint without verified acceptance in this check; A denotes verified acceptance. Published works are identified by venue. ``Not specified'' means not stated in the cited locations, not that repetitions or resource limits did not exist. Adaptive attempts, validation folds, candidate samples, raters, and bootstrap resamples are distinguished from independent agent executions. Page numbers, where supplied, count from the first PDF page.

For AI Agents That Matter, TMLR 2025 publication was verified while extraction used the author-linked July 2024 text (arXiv:2407.01502v1); the publisher PDF was not accessible during the update. Voyager was read in the author-linked arXiv:2305.16291v2 text of 19 October 2023, with TMLR 2024 publication checked separately. Agentless and Generative Agents were read in author- or institution-hosted ACM-typeset publication copies with matching title and DOI; the files were not checked for byte-for-byte identity with publisher-hosted copies. Better Turns was read as arXiv:2609.21187v1; its acceptance at the REALM Workshop at EMNLP 2026 is confirmed by the official workshop list. Other studies added during the coverage checks use the archived proceedings PDFs.

{\scriptsize\renewcommand{\arraystretch}{1.12}
\begin{xltabular}{\linewidth}{>{\raggedright\arraybackslash}p{2.1cm}>{\raggedright\arraybackslash}p{3.1cm}>{\raggedright\arraybackslash}X>{\raggedright\arraybackslash}p{2.9cm}>{\raggedright\arraybackslash}p{1.8cm}}
\caption{Detailed evidence for thirty-five focal study designs}\label{tab:appendix}\\
\toprule
\textbf{Study (version)} & \textbf{Arms and coupling} & \textbf{Population, inclusion, censoring} & \textbf{Runs; endpoint; budget} & \textbf{Location in paper} \\
\midrule
\endfirsthead
\multicolumn{5}{l}{\textit{Table~\ref{tab:appendix} (continued)}}\\
\toprule
\textbf{Study (version)} & \textbf{Arms and coupling} & \textbf{Population, inclusion, censoring} & \textbf{Runs; endpoint; budget} & \textbf{Location in paper} \\
\midrule
\endhead
\bottomrule
\endlastfoot
ToolRet~\cite{toolret} (Findings ACL 2025) & Oracle vs.\ retrieved tools; pre/post-training retrievers; fixed executor within each comparison & Retrieval corpus: 7,615 queries and 43,215 tools; downstream: ToolBench G1--G3, GPT-3.5 and ToolLlama & Retrieval metrics and pass rate; repetitions/coupling not specified in \S7; no common token cap stated there & \S3, \S5, \S7, Fig.~6, \S8 \\
AWM~\cite{awm} (ICML 2025) & Workflow memory vs.\ baselines; online and offline variants & WebArena and Mind2Web; website-specific learning; cross-template and cross-domain tests & Temperature 0; per-task repetitions not specified in \S3; success, steps, token breakdown; no equal-total-token cap & \S2--4, Tables 1--4, App.~D--E \\
ASI~\cite{asi} (COLM 2025; author copy) & Vanilla, AWM, and ASI with Claude; shopping-site verification/format ablations & 812 WebArena examples; additional scaled-up activities and transfer tests & Task success, steps, and checkpoint completion in separate experiments; repetitions/coupling not specified in \S3; macro-actions differ from primitive steps & \S2.3, \S3, Tables 1, 3; \S4, Table 4 \\
AI Agents That Matter~\cite{agentsmatter} (TMLR 2025; author-linked v1) & Agent architectures vs.\ retry, warming, and escalation; DSPy optimization & 164 HumanEval tasks in modified benchmark; HotPotQA retrieval evaluation & Five runs per configuration in Figs.~1--2; accuracy and dollar cost, with fixed/variable cost distinction; no skill-load intervention & \S2--4, Figs.~1--2, App.~A--B (author-linked text) \\
StableTool\-Bench \cite{stabletoolbench} (Findings ACL 2024) & Cache/simulator and evaluator variants; CoT/DFS agents & 765 tasks judged solvable from 1,100 original tasks & Main Table 4: one execution per model, three evaluations; solvable pass/win rates; inference and evaluator variations distinguished & \S2--4, Tables 3--6, Fig.~7 \\
Skill Following~\cite{skillfollowing} (v1) [A] & Search tool defined vs.\ removed from the prompt; same task prompt, generation seed, and decoding in both arms & 80-task MBPP+ sample (seed 42), HumanEval+, 80-task Math500 partitions; 17 LLMs; RAE restricted to tasks where the enabled run invoked retrieval and a skill was returned & One skill-enabled and one skill-disabled execution per task in each cell; counts pooled over reruns with identical configurations where available; harmful-transition annotation and skill-content controls over MBPP+ partition seeds 42--44; correctness; no budget constraint & \S3.1--3.2 (Eq.~4), \S4.1, \S4.3, \S5.2, \S5.5, App.~A, Table 17, Limitations \\
Regression Tax~\cite{regressiontax} (v1) [P] & No skills vs.\ three skill libraries; only the library changes within a stack & 486 tasks (OfficeQA-Pro, SpreadsheetBench) $\times$ 3 model--harness stacks $\times$ 4 conditions; mechanism labels from traces by one author & One run per task and condition; variance not estimated; grader pass/fail; no budget constraint & \S3.4--3.6, \S5, \S6.2, App.~C \\
Budget study~\cite{worththeirtokens} (v2) [A] & AWM, ASI, ReasoningBank (10-step actor) vs.\ vanilla actor with 15 steps and accessibility-tree pruning & Four WebArena domains, three models; WorkArena-L1 with Qwen 3.6-27B & Three runs per domain (three seeds per task type on WorkArena-L1); success and tokens; approximate budget via step cap & \S3.2--3.3, \S4, \S5.1, \S5.3, App.~B--D \\
Skill shadowing~\cite{shadowing} (v2) [P] & Library of oracle skills vs.\ expanded libraries of 52, 102, 202 skills & SkillsBench task--model pairs kept only if every skill in the authored bundle raised pass rate by at least 4 percentage points over no skills: 38 pairs (21 Haiku 4.5, 17 Sonnet 4.6); the version examined does not state whether filtering and estimation used separate runs & 2{,}545 trajectories reported in \S4 and App.~A.1; the condition counts in Table 2 sum to 2{,}576, an unresolved discrepancy in the cited version; two-stage clustered bootstrap with 2{,}000 resamples (resamples, not runs); pass rate; pooled drop at 202 skills 0.21 (21 percentage points), largest single-model drop 0.26 (Haiku 4.5) & \S3 (Assumption~1), \S4, App.~A.1, Tables 1, 4, 5 \\
SkillsBench \cite{skillsbench} (v4) [A] & No skills vs.\ curated per-task bundle; self-generated skills on three configurations & 87 tasks, 18 model--harness configurations & Three trials per cell, fresh container per run; task-macro pass rate and normalized gain; no budget constraint & \S3--4, \S5.1, Table 2, Fig.~5, App.~D.6, App.~N \\
SkillApt~\cite{skillapt} (v1) [P] & WITH vs.\ WITHOUT a candidate skill; task, model, decoding, environment, evaluator shared & Frozen SRA-Bench split: 113 confirmatory states, 111 complete after two no-skill timeouts were censored; infrastructure failures censor the pair & One WITH and one WITHOUT execution per candidate--state pair; uncertainty by paired bootstrap (resamples, not runs); correctness first, cost as tie-break & \S3.1--3.3, \S5.2, App.~I \\
RADEG~\cite{radeg} (v1) [P] & Gate predicts non-zero verifier reward from query--bundle features; separate perturbation study compares bundle compositions & 72 queries with bundles and one-skill perturbations; 288 rollouts; query-level held-out splits & Each pair executed once; non-zero verifier reward; fixed call budgets in gate evaluation & \S3, \S4.1, \S4.4, \S5, App.~A.2, App.~B \\
Confidence-gated retrieval~\cite{confgated} (v1) [P] & Raw vs.\ calibrated confidence-to-action mappings on the same fixed paths & HotpotQA and MuSiQue; Mistral, GPT, Qwen models; calibration map fitted on a separate split & Traces generated once at temperature 0 and replayed without new generation; source-example counts per split in App.~A.4, Table 5; accuracy, coverage, retrieval use; fixed budget & \S3.2, \S4.1, \S5, \S7, App.~A.2, A.4 \\
Replay Gap~\cite{replaygap} (v1) [A] & Model swap vs.\ same-model control fork at controlled points of live SWE-bench trajectories & Six run pairs, each a 30-instance sweep in one swap direction; about 900 rollouts are branch continuations, not independent tasks & Temperature 0; action edit distance and resolution; step budgets & \S3, \S4.1--4.5, Tables 1, 3, \S6 \\
Failure prevention~\cite{vasudev2026} (v1) [P] & Agent vs.\ agent with critic intervention on the same tasks & HotPotQA (100 tasks), GAIA (30), ALFWorld (202); several models; 50-task pilot & Two or three seeds, depending on configuration; App.~A lists seeds 42, 123, and 456 for the configurations tabulated there; success, recovery and disruption counts; 15-action cap and at most three critic interventions per episode, no matched total token or cost budget reported & \S3 (setup), \S4, Table 3, \S7, App.~A (Tables 11--18) \\
State-Path menu~\cite{menu} (v1) [A] & Learned menu vs.\ baseline menu constructors; agent unchanged & ToolBench online tasks; other evaluators, executors, and libraries for transfer & One deterministic run per task for the main result (one executor); 8-call budget; online success & \S4.1--4.2, App.~A.2 \\
Capability Pages~\cite{capabilitypages} (v1) [P] & Skill cards with vs.\ without negative-boundary text; same executor in both arms & SRA-Bench (26{,}262 skills, 5{,}400 questions, six datasets); four executors & Five solving rounds per executor--dataset cell, same sampling protocol in both arms; end-to-end task success; no budget constraint & \S3.4, \S5.3, Table 5 \\
TabAgent~\cite{tabagent} (v1) [P] & LLM shortlist head vs.\ trained classifier; rest of the agent unchanged & AppWorld & Methods with stochastic behavior run five times per task with distinct seeds; runs of deterministic methods not separately reported; task success, latency, cost & \S4, \S5, App.~A \\

Iterative tool retrieval~\cite{iterativetoolretrieval} (Findings EMNLP 2024) & Feedback-trained iterative retriever vs.\ ToolRetriever; shared ToolLLaMA executor & TR-bench ranking tests; downstream usage comparison restricted to ToolBench I2 & Several retriever training runs, middle result reported; downstream repetition/coupling not specified in \S5.1--5.2. NDCG, pass/win rates, latency; ten candidates, three iterations, no common token cap stated & \S5.1--5.4, Tables 2--7; pp.~5--7 \\
Chameleon \cite{chameleon} (NeurIPS 2023) & Full generated programs vs.\ selected modules disabled; ChatGPT & ScienceQA and TabMWP; disabling analysis specifies 500 test examples & Temperature 0; independent repeats/coupling not specified for Table 5. Answer accuracy; planner 128-token limit, default module 512-token limit, no total cap stated & \S5.2, Table 5; App.~A.2; pp.~9, 15--16 \\
HuggingGPT \cite{hugginggpt} (NeurIPS 2023) & Alternative controllers; planning, selection and response stages rated & 130 constructed requests from task combinations; combinations unable to yield new requests discarded & Judgments averaged across three raters, not repeated executions; temperature 0, repeats/coupling not specified. Passing, rationality and final success; no common resource cap stated & \S4.1, 4.5, Table 8; App.~A.1.6; pp.~7, 10, 15 \\
CodeAct~\cite{codeact} (ICML 2024) & Python vs.\ JSON/text actions within each model; before agent fine-tuning & API-Bank level-1 atomic calls; 82 M3ToolEval instances with authored tools & Repeats/coupling not specified in \S2.2--2.3. Atomic output correctness; multi-turn exact-match success and mean turns. Stops at success or ten turns; calls per turn can differ & \S2.2--2.3, Tables 2--3; pp.~4--5 \\
Agentless~\cite{agentless} (FSE 2025; author copy) & Fixed repair pipeline vs.\ leaderboard/reported agents; heterogeneous models/harnesses & Main comparison: 300 SWE-bench Lite issues; later filtered-set analysis separate & Forty patches and forty reproduction-test samples per issue, not full-system repeats; repetition/coupling not specified in \S4. Resolved issues, tokens/cost; budgets vary across baselines & \S3--5, Table 1; pp.~6--12 \\
LATM~\cite{latm} (ICLR 2024) & GPT-4-made tools used by GPT-3.5/GPT-4 vs.\ CoT/direct few-shot; maker-model ablation & Six reasoning tasks; 3 training, 3 validation and 240 test instances per task & Five maker trials per model/task, not downstream repeats; test-set repeats/coupling not specified. Accuracy and cost; three proposal/verification retries, one LLM use call; construction amortized & \S5.1--5.5, Tables 2--3; pp.~6--8 \\
Reflexion~\cite{reflexion} (NeurIPS 2023) & ReAct with reflective memory vs.\ restart without reflection; same few-shot examples & Focal ALFWorld comparison: 134 environments across six task types & Twelve adaptive trials, not independent learning-curve repeats; seed coupling not specified. Task completion; restart after repeated-action heuristic or over thirty actions; last three reflections retained & \S4.1, Fig.~3; pp.~5--6 \\
ExpeL~\cite{expel} (AAAI 2024) & Insights plus retrieved trajectories vs.\ ReAct/Act; retrieval-only and insights-only ablations & Training experiences separated from held-out tasks by four-fold validation; HotpotQA, ALFWorld, WebShop; FEVER transfer & Mean/standard error over folds, not execution repeats; greedy decoding, seed coupling not specified. Task success and auxiliary reward; no matched preparation-plus-evaluation budget stated & \S4--5.4, Fig.~5; pp.~3--8 \\
Voyager~\cite{voyager} (TMLR 2024; author v2) & Library ablations; transfer with/without library in Voyager and AutoGPT & Minecraft exploration; four unseen transfer tasks in new world with inventory reset & Three trials; seed coupling not specified. Items/tech-tree progress and task completion; up to 160 exploration or 50 transfer prompting iterations, not a token cap & \S3.1--3.4, Tables 1--2; pp.~6--9 \\
Generative Agents~\cite{generativeagents} (UIST 2023; author copy) & Full memory vs.\ three nested access ablations and human responses & Interviews after two simulated days; common accumulated history, restricted by each ablation & One hundred within-subject human raters, not simulation repeats. Believability ranks/TrueSkill; no common inference budget stated; histories not regenerated by ablated agents & \S6.1--6.4; pp.~13--14 \\
LLMCompiler \cite{llmcompiler} (ICML 2024) & Parallel dependency execution vs.\ ReAct, modified ReAct and OpenAI parallel calling & HotpotQA comparison subset, movie recommendation, ParallelQA; other scenarios separate & Three runs/mean accuracy in App.~D; seed coupling not stated. Accuracy, latency, token cost; modified ReAct latency control for HotpotQA/movies, ReAct for ParallelQA; no common spending cap stated & \S5.1--5.2, Tables 1--2; App.~D; pp.~5--7, 15--16 \\
$\tau$-bench~\cite{taubench} (ICLR 2025) & Models and function-calling/ReAct/Act strategies with simulated users & Authored retail/airline tasks; instructions and target database changes fixed across repetitions & At least three trials/task for main table; agent/user temperatures 0/1. Database/response success and repeated-success reliability; thirty-action limit, shared coupling not stated & \S3--5.1, Table 2, Fig.~4; pp.~3--7 \\
ToolEmu~\cite{toolemu} (ICLR 2024) & Standard vs.\ adversarial emulator with same agent; separate agent/prompt comparisons & 144 curated risk cases; validation subset of 100 cases yields 200 trajectories & One trajectory per emulator per validation case; separate three-run variance check, not a universal cell count. Safety/helpfulness and failures; no common resource cap stated & \S4--5, Tables 2--4 and footnote; pp.~7--9 \\
InjecAgent~\cite{injecagent} (Findings ACL 2024) & Agents and base/enhanced attacks from assumed correct initial tool call & 1,054 cases per attack setting; ASR-valid excludes invalid outputs, ASR-all includes them & Temperature 0; independent repeats/coupling not specified. Next harmful call or extraction/exfiltration sequence, not a full workflow; no common token cap stated & \S2.2--3.1, App.~B, C.1; pp.~3--6, 14--16 \\
AgentDojo \cite{agentdojo} (NeurIPS 2024) & Agents, attacks and defenses; benign utility separately measured & 97 user tasks; 629 task/attack-target cases across four environments; aggregation weights matter & Confidence intervals, but execution repeat count/coupling not specified in cited setup. Benign/attacked utility and targeted attack success; no common cross-agent resource cap stated & \S3--4, App.~C--D; pp.~3--9, 20--21 \\
Re-Invoke \cite{reinvoke} (Findings EMNLP 2024) & Re-Invoke vs.\ trained ToolLLM retriever and supplied reference tools; fixed ToolLLaMA+DFSDT & Six ToolBench subsets spanning I1--I3; separate ranking tests on ToolBench and ToolE & Reproduced pass rates, not a stated repetition count; seed coupling not specified. GPT-3.5-turbo grading; limited execution budgets, numeric cap not restated; offline indexing and extra online LLM call & \S4--5.3, Table 2; \S6.4; pp.~5--7, 9 \\
Better Turns \cite{betterturns} (v1) [A] & Four model pairs before/after SFT; gold-history scoring vs.\ model-history replay with fixed user turns & 84 held-out conversations, 542 next-action examples; workflow outcomes on 77 requiring tools; exact reference-call matching & Repeats, coupling and total resource cap not specified. Best deterministic completion 8/77; holistic success 0/77 for every configuration. Up to two retries on invalid tool calls & \S3--4, Table 1; Limitations; pp.~2--5 \\
\end{xltabular}}

\subsection{Targeted Readings by Topic}\label{app:coverage}

Table~\ref{tab:coverage_sources} records the remaining thirty-five targeted readings. They support the definitions of systems, populations, endpoints, and operating assumptions; they are not presented as complete design extractions. Page numbers count from the first PDF page. ALFWorld was read in the ICLR 2021 accepted manuscript deposited as arXiv:2010.03768v2; CREATOR uses the corrected ACL Anthology PDF (v2). Other entries use the archived proceedings or publisher PDFs. The interpretive boundaries reflect this review's analysis of the reported designs.

{\footnotesize\setlength{\tabcolsep}{4pt}\renewcommand{\arraystretch}{1.15}
\begin{xltabular}{\linewidth}{>{\raggedright\arraybackslash}p{2.6cm}>{\raggedright\arraybackslash}p{3.0cm}>{\raggedright\arraybackslash}X}
\caption{Targeted reading register for thirty-five further studies}\label{tab:coverage_sources}\\
\toprule
\textbf{Study} & \textbf{Source locations} & \textbf{Use in the synthesis and interpretive boundary} \\
\midrule\endfirsthead
\multicolumn{3}{l}{\textit{Table~\ref{tab:coverage_sources} (continued)}}\\
\toprule
\textbf{Study} & \textbf{Source locations} & \textbf{Use in the synthesis and interpretive boundary} \\
\midrule\endhead
\bottomrule\endfoot
\multicolumn{3}{l}{\textit{Evaluation settings}}\\*
AppWorld \cite{appworld} & \S2.1--2.2, 3.2; pp.~3--6 & Simulated app APIs and state-based goal and collateral-change tests define the environment and outcomes used by TabAgent. This benchmark definition is distinct from the effect of replacing TabAgent's shortlist head. \\
SWE-bench \cite{swebench} & \S2.1--2.2; pp.~2--3 & Repository issue/patch tasks and execution tests define resolved-issue outcomes. The benchmark source does not itself support Replay Gap's branching-rollout findings or comparisons between repair agents. \\
WebArena~\cite{webarena} & \S2.2, 3.2; pp.~3--4, 6 & Executable web environments and task success checks; the score concerns trajectories in these environments. \\
Mind2Web~\cite{mind2web} & \S2.1, 4.1--4.2; pp.~4, 7 & Evaluation supplies reference history at each step; task aggregation differs from autonomous closed-loop success. \\
WorkArena~\cite{workarena} & \S3.1, 5.2; pp.~3--4, 7 & ServiceNow task templates and programmatic validation define a domain-specific population and outcome. \\
OSWorld~\cite{osworld} & \S3.1--3.2; pp.~5--6 & Task initialization and execution-based evaluation define outcomes in interactive computer environments. \\
AgentBench~\cite{agentbench} & \S2--3; pp.~3--5 & Multiple environments broaden the tested population; breadth alone does not isolate a component effect. \\
API-Bank~\cite{apibank} & \S2.1, Figs.~1--2; pp.~2--3 & API calling, retrieval, and planning are evaluated at distinct levels. \\
BFCL~\cite{bfcl} & \S4.1, 4.4; pp.~4--5 & AST-based checks differ from state and response checks in multi-turn evaluation. \\
ToolSandbox~\cite{toolsandbox} & \S2.3, 3, App.~A.7; pp.~4--5, 16 & Stateful execution, ordered milestones, and forbidden events distinguish intermediate behavior from final success. \\
AgentBoard~\cite{agentboard} & \S2.2, 3.2; pp.~4--6 & State matching and annotated subgoals support progress measures alongside final success. \\
ToolQA \cite{toolqa} & Sections 3.1--3.3, PDF pp. 4--5, Tables 1--2; Sections 5.1--5.4, PDF p. 8 and Figure 3. & Questions are designed around external reference corpora and tool access; error analysis separates arguments, source selection, and hallucinated observations. These benchmark diagnostics do not establish causal stage contributions or eliminate memorization for every future model. \\
WebShop \cite{webshop} & Section 3.1, PDF pp. 4--5, Equation (1), Evaluation metrics paragraph; Section 3.2, PDF p. 5. & A controlled shopping simulator distinguishes graded purchase reward from complete instruction satisfaction. Its two endpoints show why reporting a reward gain alone does not establish an equal gain in fully successful tasks. \\
ALFWorld \cite{alfworld} & Section 2, PDF p. 3, high-level action list and seen/unseen split; Section 4.2 and Table 2, PDF p. 6. & Aligned household tasks connect abstract textual policies with embodied execution. High-level actions expand into several physical actions, and scene/domain shifts affect transfer, limiting comparisons that equate action counts or success across the two interfaces. \\
ScienceWorld \cite{scienceworld} & Section 4, PDF p. 5 (printed p. 11283), Tasks/Train-Development-Test/Goals and Rewards; Appendix B.1, PDF p. 16 (printed p. 11294). & Parametric science tasks award normalized credit for required and optional subgoals. The resulting score captures task progress as well as completion, so it should not be interpreted as a binary success probability. \\
GAIA \cite{gaia} & Sections 3.1--3.3, PDF pp. 4--6, especially Evaluation (pp. 5--6) and Capabilities coverage (p. 6). & General-assistant questions combine browsing, files and other tools but are scored by normalized final-answer matching. Multiple solution paths are possible, so a correct answer does not by itself verify a particular tool invocation or web-state change. \\
AgentGym \cite{agentgym} & Sections 4.1--4.3, PDF pp. 4--5 (printed pp. 27917--27918), Table 2 and Benchmark construction paragraph. & A unified framework combines diverse environments with distinct sampling rules and success or reward endpoints. It supports broader evaluation, while any pooled comparison still depends on the chosen environment weights and interpretation of those endpoints. \\
AndroidWorld \cite{androidworld} & Sections 3.2--3.4, PDF pp. 5--7, Table 2 on p. 6; Section 4.2, PDF p. 8. & Fixed application and OS versions support reproducibility while seeded task parameters vary goals and starting states. Device-state checks assess outcomes; the sampled parameters and action limits remain part of the evaluation population and resource protocol. \\
VisualWebArena \cite{visualwebarena} & Sections 3--3.3, PDF pp. 3--5 (printed pp. 883--885); evaluation primitives and Table 2 on PDF p. 6 (printed p. 886). & Visually grounded web tasks use task-specific binary validators, including text, visual and semantic checks. A common success label therefore need not imply an identical measurement rule, and some validators depend on learned judging models. \\
SWE-agent \cite{sweagent} & Section 2, PDF p. 3; Section 3, PDF pp. 3--4; Section 4 Metrics paragraph, PDF p. 5; Section 5.1, PDF pp. 5--7. & Interface design changes commands, feedback and history while keeping the language model fixed. Budget exhaustion triggers automatic patch submission, so the evaluated treatment includes both the interface package and the rule for terminating execution. \\
AssistantBench \cite{assistantbench} & Sections 3.1--3.4, PDF pp. 3--5 (printed pp. 8940--8942); Section 5.1, PDF p. 6, execution-step cap. & Open-web information tasks report partial-credit answer scores, answer rate and accuracy among non-abstentions. Their separation exposes a denominator choice: higher conditional accuracy does not alone establish better performance across all assigned tasks. \\
WorkArena++ \cite{workarenapp} & Sections 2.2--3.1, PDF pp. 3--5; Section 4.1 and Section 4.2 Maximum number of steps, PDF p. 7. & Compositional workplace workflows appear with explicit instructions or ticket-based goals requiring knowledge-base consultation. These levels change information availability, while seeded curricula and step limits further define the comparison; a gap cannot isolate retrieval alone. \\
\multicolumn{3}{l}{\textit{Tool construction and training}}\\*
ToolkenGPT~\cite{toolkengpt} & \S3.1--3.2; pp.~4--5 & Learned tool-token embeddings change calling behavior while the base model remains frozen. \\
CRAFT~\cite{craft} & \S2.1--2.2; pp.~3--5 & Offline creation, verification, and deduplication precede retrieval at inference. \\
CREATOR~\cite{creator} & \S3.1--3.4; pp.~3--5 & Creation, decision, execution, and rectification form a package of interventions. \\
AnyTool \cite{anytool} & Section 3.2, PDF p. 4, Figure 4 and Equations (1)--(2); Sections 4.1--4.2, PDF p. 5; Sections 5.2--5.3, PDF p. 7, Tables 2--3. & Hierarchical retrieval and self-reflection are evaluated with a revised pass-rate rule. Queries are manually retained when judged solvable using the API pool, so reported success concerns this filtered population and the chosen solution evaluator. \\
ToolGen \cite{toolgen} & Sections 3.2--3.6, PDF pp. 4--5; Section 4.4 and Table 3, PDF p. 8; Sections 5.1--5.2 and Table 4, PDF p. 9. & Tool tokens integrate retrieval and calling through staged training. Ranking metrics and StableToolBench outcomes are reported separately; vocabulary, training, action generation, and benchmark execution rules make the intervention broader than an external retriever replacement. \\
ToolACE \cite{toolace} & Section 2.1 and Figure 1, PDF p. 3; Sections 2.2.3--2.3, PDF pp. 5--6; Sections 3.1--3.3.2 and Figure 3, PDF pp. 6--8. & Synthetic API definitions and dialogues are checked before function-calling training and BFCL/API-Bank evaluation. Validation ablations also change retained dataset size, and synthetic-data checks do not directly measure live API execution success. \\
Agent Lumos \cite{agentlumos} & Sections 2.1--2.3, printed p. 12382 (PDF p. 3); Sections 3.2--3.3, printed p. 12384 (PDF p. 5); Table 3 and Limitations, printed p. 12388 (PDF p. 9). & Planning, grounding, and execution are modularized, with one-pass and iterative training formulations compared across tasks. The intervention changes the trained agent architecture; it does not isolate supplying a skill to a fixed inference-time agent. \\
ToolPlanner \cite{toolplanner} & Sections 4.1--4.3.2, printed pp. 18318--18319 (PDF pp. 4--5), Figure 4 and Equation (1); Section 5.5 and Tables 5--6, printed p. 18322 (PDF p. 8). & Candidate tags and planned paths guide tool use, with separate matching and task-completion feedback. Matching tests compliance with instruction-specified tool labels; it does not establish that those tools are necessary or causally improve the outcome. \\
\multicolumn{3}{l}{\textit{Memory and search}}\\*
Self-Refine \cite{selfrefine} & Sections 2 and 3.1, PDF pp. 3--5; Algorithm 1; Section 3.2, PDF pp. 5--6, for task-specific evaluation metrics. & Uses a single model for iterative feedback and revision, retaining history within an example. Evaluation depends on task-specific metrics and stopping rules; comparisons with one-step generation also change the allocation of inference calls. \\
Tree of Thoughts \cite{tot} & Section 3, PDF pp. 3--4 (thought generation/evaluation and search); Section 4.1, PDF p. 5 (Game of 24 breadth and repeated value samples). & Generates and evaluates intermediate reasoning states with bounded tree search. Branching, depth, and repeated evaluations affect inference allocation; task success on the selected reasoning problems does not directly measure a retrieved skill component. \\
Language Agent Tree Search (LATS) \cite{lats} & Section 4, PDF pp. 4--6 and Figure 2; limitations, PDF p. 9, on additional inference cost and restoration of earlier environment states. & Combines tree search, environment feedback, and reflections on failed trajectories. Search budgets and the ability to restore prior environment states are part of its operating conditions; full-system comparisons alone do not isolate individual mechanisms. \\
A-Mem \cite{amem} & Sections 3.2--3.4, PDF pp. 4--5; Section 4.1, PDF pp. 5--6, for datasets and evaluation metrics. & Constructs linked memory notes, updates their attributes, and retrieves them for questions about long conversations. Main metrics assess answer quality on LoCoMo and DialSim; these outcomes do not directly establish autonomous tool-trajectory success. \\
\multicolumn{3}{l}{\textit{Risk endpoints}}\\*
AgentHarm \cite{agentharm} & Sections 3.1.2 and 3.1.3, PDF p. 5 (synthetic tools, task grading, refusal); Section 3.2, PDF p. 6 (scope and limitations). & Scores progress and refusal on malicious multi-step tasks executed with synthetic tools. The authors explicitly treat the tasks as harm proxies without real side effects; scores therefore do not measure completed real-world harmful outcomes. \\
\end{xltabular}}

\end{document}